\documentclass[a4paper,11pt]{article}
\usepackage{cite}
\usepackage{authblk}
\usepackage{graphicx}
\usepackage{caption}
\usepackage{float}
\usepackage{commath,amssymb,mathtools,cancel}
\usepackage[margin=0.8in]{geometry}
\usepackage{setspace}
\usepackage{hyperref}
\hypersetup{
  colorlinks   = true, 
  urlcolor     = blue, 
  linkcolor    = black, 
  citecolor    = red 
}
\numberwithin{equation}{section}

\title{Bulk-to-bulk photon propagator in AdS}

\author[1]{Radu N. Moga\thanks{R-N.Moga@soton.ac.uk}}
\author[1]{Kostas Skenderis\thanks{K.Skenderis@soton.ac.uk}}

\affil[1]{STAG Research Centre \& Mathematical Sciences, University of Southampton, Highfield, SO17 1BJ Southampton, UK}

\date{}

\DeclareMathOperator{\sech}{sech}
\DeclareMathOperator{\csch}{csch}

\begin{document}

\maketitle

\begin{abstract}
    We study the photon bulk-to-bulk propagator in AdS in various gauges, including axial, Coulomb, and the standard covariant gauge. We compute the propagator using both momentum and position space techniques. We ensure the propagators obtained obey the right subsidiary conditions arising from gauge invariance. In particular, BRST invariance implies a relation between the longitudinal components of the gauge field propagator and the ghost bulk-to-bulk propagator. Our method relies on decomposing the components of the propagator in terms of independent tensor structures and solving for the form factors. We recover some previously existing results and obtain new expressions for the propagator in other gauges. The propagator in axial and Coulomb gauge is simpler in momentum space, as momentum space makes manisfest the translational invariance in the boundary directions, while the position space expression is the simplest in the covariant Fried-Yennie gauge. In this gauge the propagator has an improved IR behavior, somewhat analogous to the UV improved behavior associated with the Landau gauge in flat space. The results readily extend to Yang-Mills fields.
\end{abstract}

\newpage

\tableofcontents

\section{Introduction}

An essential component of perturbation theory is the propagator of the field. It is formally defined as the inverse of the differential operator appearing in the kinetic term of the Lagrangian. In the presence of gauge invariance the kinetic operator has zero eigenmodes and to obtain the propagator we need to fix the gauge and introduce corresponding ghost fields. The general procedure is well understood for any gauge choice and the gauged-fixed action can be obtained by BRST quantization, in flat and in curved spacetimes. However, apart from the case of flat space, explicit expressions for propagators are rare. In this paper we discuss the case of gauge fields in (Euclidean) Anti-de Sitter spacetime.

One motivation for this computation comes from the AdS/CFT correspondence \cite{Maldacena:1997re,Gubser:1998bc,Witten:1998qj}. Tree-level diagrams were computed already in the foundational papers and the early AdS/CFT literature (see \cite{DHoker:2002nbb} for a review). More recently, loop-level computations of Witten diagrams have also been carried out \cite{Penedones:2010ue,Fitzpatrick:2011dm,Aharony:2016dwx,Alday:2017xua,Aprile:2017bgs,Giombi:2017hpr,Yuan:2018qva,Bertan:2018khc,Bertan:2018afl,Ghosh:2019lsx,Ponomarev:2019ofr,Carmi:2019ocp,Meltzer:2019nbs,Albayrak:2020isk,Albayrak:2020bso,Costantino:2020vdu,Carmi:2021dsn,Alday:2021ajh,Heckelbacher:2022fbx,Banados:2022nhj,Chowdhury:2023khl,Ankur:2023lum,Carmi:2024tzj,Ciccone:2024guw,Chowdhury:2024snc,Chowdhury:2025ohm} but there are only very few works involving gauge field loops \cite{Albayrak:2020bso,Ankur:2023lum,Carmi:2024tzj,Ciccone:2024guw}. Another motivation for studying gauge fields in AdS is that AdS may be considered as an IR regulator for the flat space theory \cite{Callan:1989em}.

There are many different gauge choices, though in particular two of them are usually considered in AdS. The axial gauge consists of adding a gauge fixing term of the form $\frac{1}{2\xi}\left(n^\mu A_\mu\right)^2$ to the Lagrangian, where $n^\mu=(z,\vec 0)$ in Poincare coordinates such that $n_\mu n^\mu=1$. Taking $\xi\to0$ corresponds to fixing $A_0=0$. This choice of gauge has the advantage of being very simple to work with, especially in the context of the AdS/CFT correspondence, as the relation between bulk fields and CFT sources is particularly transparent in this gauge. It also leads to (relatively) simple expressions for the propagator in momentum space. Another advantage is that the corresponding ghosts are not propagating. On the other hand, in flat space the propagator contains unphysical IR poles \cite{Leibbrandt:1987qv} and this makes the computation of loop diagrams tricky. However, AdS regulates such poles \cite{Marotta:2024sce} and it could provide a suitable framework for flat space higher-loop computations in axial gauge. Another disadvantage of this gauge is that it is not covariant. The axial gauge has often been used when computing correlators in momentum space, see for example \cite{Raju:2011mp, Albayrak:2020bso, Marotta:2024sce}.

Another choice of gauge is the standard covariant gauge \cite{Allen:1985wd} which consists of adding a gauge fixing term of the form $\frac{1}{2\xi}\left(\nabla^\mu A_\mu\right)^2$. This is the AdS-covariant version of the most standard gauge choice used in flat space (the $R_\xi$ gauge). The advantage of this gauge is that it preserves AdS covariance. This is the gauge that is used most often in position space calculations. On the other hand, for generic values of $\xi$ the expressions for the propagator in AdS are rather complicated. This gauge choice has been discussed recently in \cite{Ciccone:2024guw}.

A third, not as common, gauge choice is the Coulomb (or `boundary-transverse') gauge in Poincare coordinates for which the gauge fixing term is $\frac{1}{2\xi}\left(g^{ij}\partial_j A_i\right)$, where the sum excludes the radial coordinate. This is one of the gauges where electrodynamics in flat space was originally quantized \cite{Schwinger:1948yk}, and has the advantage that only photons with helicities $\pm 1$ are introduced, see \cite{Weinberg:1995mt} for a textbook treatment. This is not a covariant gauge, but, similarly to the axial gauge, leads to a rather simple expression for the propagator in momentum space as it exploits the translational invariance of the AdS metric along the boundary directions in these coordinates.

There is a long history regarding the gauge field propagator in (A)dS and various issues with it dating back to the 80's. In a seminal work from 1986, Allen and Jacobson \cite{Allen:1985wd} computed the vector propagator in position space in the covariant, Feynman gauge ($\xi=1$) in both dS and AdS. They started with a covariant Ansatz for the propagator and used it to solve the Green's function equation that the propagator must satisfy. They solved this equation at separated points and then in order to fix the undetermined coefficients they matched the short-distance behaviour of their answer with the short-distance behaviour of the gauge field propagator in flat space.

In the years since, there has still been a lot of discussion about the photon propagator in the covariant gauge in dS, both generalizing the early results as well as discussing various problems (and their potential resolutions), see \cite{Tsamis:2006gj,Youssef:2010dw, Frob:2013qsa, Domazet:2014bqa, Narain:2014oja, Gibbons:2014zya,Frob:2017gez,Glavan:2022pmk,Glavan:2022dwb,Glavan:2022nrd,Domazet:2024dil,Glavan:2025iuw} for a sample of such papers. Many of these papers address IR issues associated with the dS photon propagator, which are not directly relevant for our discussion. What is relevant for us is the constraints imposed due to BRST invariance discussed in \cite{Frob:2017gez,Glavan:2022nrd}: BRST invariance implies a Ward identity that connects the gauge field and ghost propagators. This provides a new check for propagators, but more importantly for us, it also yields a more tractable equation that may be used when solving for the propagators.

In AdS there has been, comparatively, much less discussion of the photon propagator. There are a number of treatments in the literature. The first is the computation by Allen and Jacobson \cite{Allen:1985wd} mentioned above for the photon propagator in the covariant, Feynman gauge ($\xi=1$), which led to rather complicated expressions for the propagator. This was followed up by Liu and Tseytlin \cite{Liu:1998ty} who set out to compute the propagator in axial gauge.\footnote{In their paper they called it the Coulomb gauge, but their starting point was the gauge condition $A_0=0$ which we refer to as the axial gauge.} They did an on-shell computation (they used current conservation, which is valid only on-shell), and as we will see in section \ref{section Coulomb gauge}, the propagator they computed is actually the Coulomb gauge propagator, {\it i.e.}\ it is the propagator that results from the gauge condition $g^{ij}\partial_j A_i=0$, where the sum excludes the radial coordinate. There is no contradiction here, as on-shell one may impose both the axial and Coulomb gauge fixing conditions at the same time, see \cite{Marotta:2024sce} and the discussion in section \ref{sec: axial_to_Coulomb}. Moreover, for the purpose of computing holographic correlators at tree-level, which only needs the on-shell action, their analysis is sufficient. However, the ghost action for the axial and Coulomb gauges is different (as we discuss later). Thus, the propagator in \cite{Liu:1998ty} would need to be accompanied by the Coulomb gauge ghost action to obtain correct results at loop-level.

Following up on this, D'Hoker and Freedman \cite{DHoker:1998bqu,DHoker:1999bve} computed the transverse part of the photon propagator in the covariant, Feynman gauge ($\xi=1$) using different variables than in \cite{Allen:1985wd}. In their variables this part turned out to be a rather simple expression, related to the propagator of a massive scalar, as in \cite{Liu:1998ty}. The missing longitudinal part of the propagator does not contribute at tree-level and hence they could use their expression for such computations. For loop computations the longitudinal piece would be needed. A different approach was pursued by Raju \cite{Raju:2011mp}, who gave an integral representation of the propagator in momentum space for the axial gauge. He also compared his result with the Liu and Tseytlin propagator and found that the two forms are equivalent. However, in his comparison he used current conservation which is valid on-shell. The two propagators are not equivalent off-shell, {\it i.e.}\ at loop order they come with different corresponding propagators for the ghost fields. A new computation of the propagator in momentum space for the axial gauge was presented in \cite{Marotta:2024sce}. The equivalence between explicit expressions (as in \cite{Marotta:2024sce}) and integral representations (as in \cite{Raju:2011mp}) will be shown in appendix \ref{appendix integral representation}. A different integral representation of the photon propagator, now in terms of eigenfunctions of the AdS Laplacian, was presented in \cite{Ankur:2023lum,Carmi:2024tzj,Ciccone:2024guw}. The methodology is based on AdS harmonic analysis that was developed in \cite{Costa:2014kfa} (see also \cite{Leonhardt:2003qu, Leonhardt:2003sn}). Using this approach, in \cite{Ciccone:2024guw} the authors computed the position space covariant propagator for any $\xi$.

The goal of the present work is to provide a definitive answer for the propagator in AdS in various gauges and to present in which circumstances they take the simplest forms. We will work with Euclidean AdS. Expressions for Lorentzian propagators may be obtained by suitable analytic continuation. We will determine the propagator using a path-integral approach, starting with momentum space. In the case of covariant gauges, we highlight a special but relatively unknown choice of $\xi$ called the Fried-Yennie gauge \cite{PhysRev.112.1391,Yennie:1961ad} and also derive the propagator in position space. We find that the form of the propagator in axial and Coulomb gauges is simple in momentum space. This reflects the fact that these gauge choices treat differently the radial direction from the boundary directions, which are the directions we Fourier transform. On the other hand, the propagator in the covariant gauge takes a form that is more natural in position space as the gauge fixing condition combines all directions of AdS.

The rest of the paper is organized as follows. Section \ref{section generalities} contains the set-up together with a general discussion of the implications of BRST invariance for the propagator. In section \ref{section momentum space} we set up the momentum space derivation and obtain the propagator in axial, Coulomb, and standard covariant gauges, respectively. The final answers have been collected at the end of each section for the readers not interested in the derivations. In section \ref{section position space} we discuss why the Fried-Yennie gauge is special in flat space and in section \ref{section position space 2} we use the intuition from flat space to derive a simple form of the propagator in AdS in this gauge. Section \ref{section conclusion} contains our conclusions. The paper contains a number of appendices. In appendix \ref{appendix flat space propagators} we review the flat space propagators in momentum space and verify the BRST relation between the gauge field propagator and the ghost propagator.  In appendix \ref{appendix integral representation} we give an integral representation for the AdS propagators in the same fashion as that of \cite{Raju:2011mp} in axial gauge, which also allows us to find the position representation of the propagators in axial and Coulomb gauges. Appendix \ref{appendix position space formulas}  contains useful formulas for position space calculations and in appendix \ref{appendix position space propagator in covariant gauge for arbitrary xi} we present the detailed form of the position space photon propagator in the covariant gauge for arbitrary values of $\xi$ and $d$. We also include as ancillary files some relevant Mathematica notebooks which contain the derivations of the propagator in the covariant gauges in both momentum and position space. The general results are valid for $d>2$ and we present the $d=2$ case in appendix \ref{app: d=2}.
\section{Generalities}\label{section generalities}

\subsection{BRST invariant Maxwell action}

Our starting point is the BRST invariant action for a free $U(1)$ gauge theory (we work in Euclidean signature). Following the conventions in \cite{Polchinski:1998rq}, we have
\begin{equation}
    Z=\int \mathcal{D}A \mathcal{D}B \mathcal{D}c \mathcal{D}b \, e^{-S}, \quad S=\int \mathrm{d}^{d+1}x \sqrt{g}\left[\frac14 F^{\mu\nu}F_{\mu\nu}-iBF(A)+b\delta_{\rm gauge}^{(c)}F(A)+\frac{\xi}{2}B^2\right],
\end{equation}
where $B$ is an auxiliary field, $F(A)$ is a gauge fixing condition of choice, $c$ and $b$ are the ghost and antighost fields, $\delta_{\rm gauge}^{(c)}F(A)$ is a gauge transformation with parameter $c$ ({\it i.e.}\ $\delta_{\rm gauge}^{(c)}A_\mu=\partial_\mu c$) of the gauge fixing condition $F(A)$, and $\xi$ is the gauge fixing parameter. The gauge fixing conditions $F(A)$ we consider in this paper are linear in $A$.

The action is invariant under the BRST transformation
\begin{align}
    \delta_\epsilon A_\mu&=-i\epsilon\partial_\mu c,\\
    \delta_\epsilon B&=0,\\
    \delta_\epsilon b&=\epsilon B,\\
    \delta_\epsilon c&=0,
\end{align}
where $\epsilon$ is an anticommuting parameter. Note that for $F$ linear in $A$, $\delta_{\rm gauge}^{(c)}F(A)=F(\partial c)$. The equation of motion (EoM) for the auxiliary field is
\begin{equation}
    B=\frac{i}{\xi}F(A),
\end{equation}
and integrating it out gives
\begin{equation}
    Z=\int \mathcal{D}A \mathcal{D}c \mathcal{D}b \, e^{-S}, \quad S=\int \mathrm{d}^{d+1}x \sqrt{g}\left[\frac14 F^{\mu\nu}F_{\mu\nu}+\frac{1}{2\xi}\left(F(A)\right)^2+bF(\partial c)\right],\label{action}
\end{equation}
where now the BRST transformation that leaves the action invariant is given by
\begin{align}
    \delta_\epsilon A_\mu&=-i\epsilon\partial_\mu c,\\
    \delta_\epsilon b&=\frac{i\epsilon}\xi F(A),\\
    \delta_\epsilon c&=0.
\end{align}

As mentioned, the action is free, so the ghosts decouple. However, they will be relevant for our derivation of the propagator. We define the gauge field and ghost propagators as the corresponding 2-point correlators,
\begin{align}
    \left\langle A_\mu(x)A_\nu(y)\right\rangle&=G_{\mu\nu}(x,y),\\
    \left\langle c(x)b(y)\right\rangle=-\left\langle b(y)c(x)\right\rangle&=G_{\rm ghost}(x,y).
\end{align}
Note that the photon propagator must obey the following exchange property
\begin{equation}
    G_{\mu\nu}(x,y)=\left\langle A_\mu(x)A_\nu(y)\right\rangle=\left\langle A_\nu(y)A_\mu(x)\right\rangle=G_{\nu\mu}(y,x).\label{exhange property}
\end{equation}

BRST invariance implies a relation between the gauge and ghost propagators. Indeed, any correlator of a BRST-exact object must vanish, and we have
\begin{equation}
    0=\left\langle\delta_\epsilon(b(x)A_\nu(y))\right\rangle=\frac{i\epsilon}{\xi}\left\langle F(A(x))A_\nu(y)\right\rangle+i\epsilon\left\langle b(x)\partial_\nu c(y)\right\rangle,
\end{equation}
which implies that
\begin{equation}
    \left\langle F(A(x))A_\nu(y)\right\rangle=\xi\partial_{y,\nu}G_{\rm ghost}(y,x).\label{brst constraint}
\end{equation}
For a linear gauge fixing condition $F$, the left-hand side is just some tensor contraction of the photon propagator. So even if the ghosts decouple from the theory, BRST invariance constrains the ghost and gauge field propagators. For example, in the case of covariant gauges, {\it i.e.}\ $F(A)=\nabla^\mu A_\mu$, the above equation becomes
\begin{equation}
    \nabla^\mu G_{\mu\nu}(x,y)=\xi\partial_{y,\nu}G_{\rm ghost}(y,x).
\end{equation}
This relation is instrumental in proving perturbative unitarity in flat space: it shows that the longitudinal component of the photon propagators cancels against the ghost propagators. This relation and its use in the context of the photon propagator in de Sitter were discussed in \cite{Glavan:2022nrd}.

\subsection{Relation between the propagators' and BRST constraint equations}

Before moving on to the specific case of EAdS, it is worth noting some remarks on the equations that determine the ghost and gauge field propagators and the BRST constraint equation that relates them. For concreteness, we will stick to the case of standard covariant gauges. In this case, we have
\begin{align}
    \nabla^\mu\nabla_\mu G_{\rm ghost}(x, y)&=\frac{1}{\sqrt g}\delta^{(d+1)}(x-y),\label{gh eq int}\\
    \left(g^{\mu\nu}\nabla^\sigma\nabla_\sigma-\nabla^\mu\nabla^\nu+\frac{1}{\xi}\nabla^\nu\nabla^\mu\right)G_{\mu\rho'}(x, y)&=-\frac{1}{\sqrt g}\delta^\nu_{\rho'}\delta^{(d+1)}(x-y),\label{gauge eq int}\\
    \nabla^\mu G_{\mu\nu'}(x, y)&=\xi\nabla_{\nu'}G_{\rm ghost}(x, y).\label{BRST eq int}
\end{align}
Here, as well as in sections \ref{section position space}, \ref{section position space 2}, and appendices \ref{appendix position space formulas} and \ref{appendix position space propagator in covariant gauge for arbitrary xi} where we work in position space, we use unprimed and primed indices to emphasize that propagators are bitensors and hence depend on two spacetime points. Unprimed indices are associated with $x$ while primed indices are associated with $y$. Importantly, we use the fact that covariant derivatives with respect to different spacetime points (so, in what follows, with unprimed and primed indices) commute. More details on how to manipulate bitensors can be found in the original works by Allen and Jacobson, for example in \cite{Allen:1985wd}. We are also assuming here that the propagators depend only on an isometry-invariant distance between the two points, and hence there is no distinction between $G_{\rm ghost}(x,y)$ and $G_{\rm ghost}(y,x)$ or $G_{\mu\nu}(x,y)$ and $G_{\mu\nu}(y,x)$, and we will drop the propagators' arguments for brevity. This will indeed be the case for EAdS as we will prove in section \ref{section position space 2}. We emphasise, however, that in general only  (\ref{exhange property}) must hold.

Since there are two unknowns, namely the photon and ghost propagators, and three equations it would seem that the system is overdetermined unless one of the equations is implied by the other two. Also, for the case of dS, in \cite{Glavan:2022nrd} the authors showed that there are solutions to the ghost and gauge field propagators' equations that do not satisfy the BRST constraint. We will now study what are the implications for the ghost propagator that come by combining the equation for the gauge field propagator with the BRST constraint equation. We start by taking the divergence of the BRST constraint equation with respect to the primed coordinate,
\begin{equation}
    \nabla^{\nu'}\nabla^\mu G_{\mu\nu'}=\xi\nabla^{\nu'}\nabla_{\nu'}G_{\rm ghost}.\label{useful}
\end{equation}
This equation will be useful later. We now consider the photon propagator equation and rewrite it by commuting the covariant derivatives and introducing the curvature tensor in the process,
\begin{equation}
    \left(g^{\mu\nu}\nabla^\sigma\nabla_\sigma+R^{\mu\nu}\right)G_{\mu\rho'}+\left(\frac{1}{\xi}-1\right)\nabla^\nu\nabla^\mu G_{\mu\rho'}=-\frac{1}{\sqrt g}\delta^\nu_{\rho'}\delta^{(d+1)}(x-y).
\end{equation}
We now take the divergence of this equation with respect to the primed coordinate,
\begin{equation}
   \left(g^{\mu\nu}\nabla^\sigma\nabla_\sigma+R^{\mu\nu}\right)\nabla^{\rho'}G_{\mu\rho'}+\left(\frac{1}{\xi}-1\right)\nabla^\nu\nabla^{\rho'}\nabla^\mu G_{\mu\rho'}=-\frac{1}{\sqrt g}\delta^\nu_{\rho'}\nabla^{\rho'}\delta^{(d+1)}(x-y),
\end{equation}
and now using equations (\ref{BRST eq int}) and (\ref{useful}) we get,
\begin{equation}
    \xi\left(g^{\mu\nu}\nabla^\sigma\nabla_\sigma+R^{\mu\nu}\right)\nabla_{\mu}G_{\rm ghost}+\left(1-\xi\right)\nabla^\nu\nabla^{\mu'}\nabla_{\mu'}G_{\rm ghost}=-\frac{1}{\sqrt g}\delta^\nu_{\rho'}\nabla^{\rho'}\delta^{(d+1)}(x-y).
\end{equation}
Commuting the covariant derivatives in the first term and simplifying, we get
\begin{equation}
    \xi\nabla^\nu\left(\nabla^\sigma\nabla_\sigma G_{\rm ghost}-\nabla^{\mu'}\nabla_{\mu'}G_{\rm ghost}\right)+\nabla^\nu\nabla^{\mu'}\nabla_{\mu'}G_{\rm ghost}=-\frac{1}{\sqrt g}\delta^\nu_{\rho'}\nabla^{\rho'}\delta^{(d+1)}(x-y).
\end{equation}
Using the fact that the ghost propagator and the delta-function are symmetric in their arguments, the above becomes
\begin{equation}\label{eq_der_gh}
    \nabla^\nu\nabla^{\mu}\nabla_{\mu}G_{\rm ghost}=\frac{1}{\sqrt g}\nabla^{\nu}\delta^{(d+1)}(x-y).
\end{equation}
This is nothing but the derivative of the ghost propagator equation! Thus, the divergence of the equation for the gauge field propagator (w.r.t.\ the second point) and the divergence of the BRST constraint equation (w.r.t.\ the second point) imply the derivative of the equation for the ghost propagator.\footnote{This means, in particular, that this consistency condition would also be satisfied for 2-point functions that fail to satisfy the Green's function equation by a constant. This is relevant for gauge theories on compact spaces.}

The important point to retain from this discussion is that all three equations for the two propagators and the BRST constraint equation provide non-trivial information. Unless all of them are checked when solutions are found for the propagators, there might be erroneous solutions. It is to be expected, however, that if for example one solves the ghost and gauge field propagators' equations and imposes the right conditions ({\it e.g.}\ boundary conditions, flat space limit, etc.) such that the solutions obtained are unique, they will automatically satisfy the BRST constraint equation as well. This can also be used the other way around, as sometimes it may not be clear {\it a priori} what the right conditions are. In such cases, the BRST constraint equation can be used to narrow down the allowed solutions of the ghost and gauge field propagators' equations to the ones compatible with BRST invariance.

We note here that up to this point the discussion has been completely general and applies to any spacetime. In appendix \ref{appendix flat space propagators} we show explicitly that the well-known flat space propagators in various gauges satisfy the BRST constraint equation. We now move on to our case of interest, namely EAdS.
\section{Momentum space propagators}\label{section momentum space}

\subsection{Euclidean AdS$_{d+1}$ in momentum space}

We now specialize to EAdS$_{d+1}$ in Poincare coordinates $x=(z,\vec x)$ with the line element given by
\begin{equation}
    ds^2=\frac{dz^2+d\vec x^2}{z^2}.
\end{equation}
We will work in momentum space along the boundary $\vec x$-coordinates and define
\begin{align}
    G_{\mu\nu}(x,y)&=\int\frac{\mathrm{d}^dp}{(2\pi)^d}\tilde{G}_{\mu\nu}(z,z',\vec p)e^{-i\vec p\cdot(\vec x-\vec y)},\\
    G_{\rm ghost}(x,y)&=\int\frac{\mathrm{d}^dp}{(2\pi)^d}\tilde{G}_{\rm ghost}(z,z',p)e^{-i\vec p\cdot(\vec x-\vec y)}.
\end{align}
Note that the exchange property (\ref{exhange property}) becomes
\begin{equation}
    \tilde{G}_{\mu\nu}(z,z',\vec p)=\tilde{G}_{\nu\mu}(z',z,-\vec p) \label{exhange property mom space}
\end{equation}
in momentum space. We decompose $\tilde{G}_{\mu\nu}(z,z',\vec p)$ as follows,
\begin{equation}
    \begin{cases}\label{ads decomposition}
        \tilde{G}_{ij}(z,z',\vec p)&=A \left(\delta_{ij}-\frac{p_i p_j}{p^2}\right)
        +B\frac{p_ip_j}{p^2}\\
        \tilde{G}_{i0}(z,z',\vec p)&=C_1\frac{-ip_i}{p}, \quad \tilde{G}_{0i}(z,z',\vec p)=C_2\frac{ip_i}{p}\\
        \tilde{G}_{00}(z,z',\vec p)&=D
    \end{cases},
\end{equation}
where $A,B,C_1,C_2$, and $D$ are to be determined scalar functions of $z,z'$, and $p$. 

In all cases considered, the equation for the ghost propagator is straightforward to solve and it will be shown explicitly in each case, while the equations for the different components of the photon propagator take the form\footnote{We will use primes to denote derivatives with respect to $z$ when working in momentum space.}
\begin{equation}
    \left(z^2\partial_z^2-(d-3)z\partial_z-z^2p^2\right)A=-z^{d-1}\delta(z-z'),
\end{equation}
\begin{align}\label{general system of eq w/out brst}
    f_1^{(F,\xi)}(B,C_2)&=-z^{d-1}\delta(z-z'),\\
    f_2^{(F,\xi)}(B,C_2)&=0,
\end{align}
\begin{align} 
    f_1^{(F,\xi)}(C_1,D)&=0,\\
    f_2^{(F,\xi)}(C_1,D)&=-z^{d-1}\delta(z-z'). \label{general system of eq w/out brst 2}
\end{align}
Here $f_{1,2}^{(F,\xi)}(\cdot,\cdot)$ are differential operators of $z$ acting on two variables which depend on the gauge fixing condition $F(A)$ and the gauge fixing parameter $\xi$. The equation for $A$ is the same for any gauge choice. This makes sense as it corresponds to the transverse part of the propagator and is thus the physical part. The equations for the other variables depend on the gauge choice, and the equations for $B$ and $C_2$ are exactly the same equations as those for $C_1$ and $D$, except that the delta-function, which was present in the $f_1$ equation for $B, C_2$, appears in the $f_2$ equation for $C_1, D$.

The equation for $A$ can be solved straightforwardly as it will be shown in a moment. The equations for the other variables are, in general, much harder to solve as they are coupled second-order differential equations.\footnote{This is true for example for the standard covariant gauge discussed in section \ref{section standard covariant gauge}.} In order to solve for those, we use the BRST constraint (\ref{brst constraint}), which in momentum space and in Poincare coordinates\footnote{The first equation comes from setting $\nu=i$ while the second equation comes from setting $\nu=0$ in (\ref{brst constraint}).} becomes
\begin{align}
    f_3^{(F)}(B,C_2)&=\xi p^2 \tilde{G}_{\rm ghost}(z',z,p),\\
    f_3^{(F)}(C_1,D)&=\xi p \partial_{z'} \tilde{G}_{\rm ghost}(z',z,p).
\end{align}
Here $f_3^{(F)}(\cdot,\cdot)$ is again a differential operator of $z$ acting on two variables which depends on the choice of $F(A)$. Putting everything together, we get
\begin{equation}\label{general system of eq}
    \left(z^2\partial_z^2-(d-3)z\partial_z-z^2p^2\right)A=-z^{d-1}\delta(z-z'),
\end{equation}
\begin{align}
    f_1^{(F,\xi)}(B,C_2)&=-z^{d-1}\delta(z-z'),\\
    f_2^{(F,\xi)}(B,C_2)&=0,\\
    f_3^{(F)}(B,C_2)&=\xi p^2 \tilde{G}_{\rm ghost}(z',z,p),
\end{align}
\begin{align}
    f_1^{(F,\xi)}(C_1,D)&=0,\\
    f_2^{(F,\xi)}(C_1,D)&=-z^{d-1}\delta(z-z'),\\
    f_3^{(F)}(C_1,D)&=\xi p \partial_{z'} \tilde{G}_{\rm ghost}(z',z,p).\label{general system of eq 2}
\end{align}
The general strategy for solving these differential equations will be as follows. We first solve the equation for $A$ which will be done below. For the first system of three coupled second-order differential equations, we will use substitution to get a second-order inhomogeneous differential equation for $B$ which we can solve. We then use one of the other equations to solve for $C_2$ knowing $B$. The equation for $C_2$ in terms of $B$ is algebraic. Finally, since we have a system of three equations and only two unknowns, we then check that the remaining equation is satisfied by the solutions found for $B$ and $C_2$. For solving the second system of three coupled second-order differential equations we apply the same procedure with $D$ taking the role of $B$ and $C_1$ taking the role of $C_2$.

\subsection{Gauge independent part of the propagator}

We now solve the equation for $A$, the transverse (hence, gauge independent) part of the propagator,
\begin{equation}
    \left(z^2\partial_z^2-(d-3)z\partial_z-z^2p^2\right)A=-z^{d-1}\delta(z-z').\label{eq for A}
\end{equation}
We first solve the equation at separated points, $z\neq z'$. In this case, the equation can be solved in terms of Bessel functions and we get
\begin{equation}
    A=c_1(p,z')z^{\frac{d-2}{2}} I_{\frac{d-2}{2}}(p z)+c_2(p,z')z^{\frac{d-2}{2}} K_{\frac{d-2}{2}}(p z).
\end{equation}
Imposing regularity of the solution sets
\begin{equation}
    A=\begin{cases}
        c_1(p,z')z^{\frac {d-2}2}I_{\frac{d-2}{2}}(p z), \quad z<z'\\
        c_2(p,z')z^{\frac {d-2}2}K_{\frac{d-2}{2}}(p z), \quad z>z'
    \end{cases}.
\end{equation}
To solve for the delta-function constraint, we impose continuity of the solution and discontinuity in its first derivative as obtained from integrating (\ref{eq for A}) from $z'-\epsilon$ to $z'+\epsilon$ and then taking $\epsilon\to0$. The first condition yields
\begin{equation}
    c_1(p,z)I_{\frac{d-2}{2}}(p z)=c_2(p,z)K_{\frac{d-2}{2}}(p z),
\end{equation}
while the second one yields
\begin{equation}
    c_2(p,z)\frac{\partial}{\partial z}\left(z^{\frac {d-2}2}K_{\frac{d-2}{2}}(p z)\right)-c_1(p,z)\frac{\partial}{\partial z}\left(z^{\frac {d-2}2}I_{\frac{d-2}{2}}(p z)\right)=-z^{d-3}.
\end{equation}
Using these two equations to solve for the integration constants gives
\begin{equation}
    \begin{cases}
        c_1(p,z)=z^{\frac {d-2}2}K_{\frac{d-2}{2}}(p z)\\
        c_2(p,z)=z^{\frac {d-2}2}I_{\frac{d-2}{2}}(p z)
    \end{cases},
\end{equation}
and thus
\begin{equation}
    \begin{aligned}
        A&=\begin{cases}
        (zz')^{\frac {d-2}2}K_{\frac{d-2}{2}}(p z')I_{\frac{d-2}{2}}(p z), \quad z<z'\\
        (zz')^{\frac {d-2}2}I_{\frac{d-2}{2}}(p z')K_{\frac{d-2}{2}}(p z), \quad z>z'
    \end{cases}\\
    &=\left((zz')^{\frac {d-2}2}I_{\frac{d-2}{2}}(p z')K_{\frac{d-2}{2}}(p z)\right)\Theta(z-z')+\left((zz')^{\frac {d-2}2}K_{\frac{d-2}{2}}(p z')I_{\frac{d-2}{2}}(p z)\right)\Theta(z'-z).\label{sol for A}
    \end{aligned}
\end{equation}
In what follows we make specific choices for the gauge and solve for the other (gauge dependent) components of the propagator.

\subsection{Propagator in axial gauge}\label{section axial gauge}

Choosing $F(A)=n^\mu A_\mu$, the action (\ref{action}) becomes
\begin{equation}
    S=\int \mathrm{d}^{d+1}x \sqrt{g}\left[\frac14 F^{\mu\nu}F_{\mu\nu}+\frac{1}{2\xi}\left(n^\mu A_\mu\right)^2+b n^\mu\partial_\mu c\right].
\end{equation}
We choose $n^\mu=(z,\vec 0)$. The normalization is unimportant as at the end of the calculation we will take $\xi\to0$ to get the axial gauge propagator and this sets $A_0=0$ regardless, but we note that with this choice we have $n_\mu n^\mu=g_{\mu\nu}n^\mu n^\nu=1$.

\subsubsection{Ghost propagator}

In this case, the ghost propagator is very simple. It satisfies
\begin{align}
    n^\mu\partial_\mu G_{\rm ghost}(x,y)&=\frac{1}{\sqrt g}\delta^{(d+1)}(x-y),\\
    \partial_0 G_{\rm ghost}(x,y)&=z^d\delta^{(d+1)}(x-y),
\end{align}
or, in momentum space,
\begin{equation}
    \partial_z \tilde G_{\rm ghost}(z,z',p)=z^d\delta(z-z').
\end{equation}
For $z\neq z'$, we get that $\tilde G_{\rm ghost}(z,z',p)$ is independent of $z$, so we have
\begin{equation}
    \tilde G_{\rm ghost}(z,z',p)=\begin{cases}
        c_1(p,z'), \quad z>z'\\
        c_2(p,z'), \quad z<z'
    \end{cases}.
\end{equation}
However, $\tilde G_{\rm ghost}(z,z',p)$ must vanish at the boundary, {\it i.e.}\ as $z\to0$, so $c_2(p,z')=0$. Thus, we get
\begin{equation}
    \tilde G_{\rm ghost}(z,z',p)=c_1(p,z')\Theta(z-z')
\end{equation}
and
\begin{equation}
    \partial_z \tilde G_{\rm ghost}(z,z',p)=c_1(p,z')\delta(z-z'),
\end{equation}
so $c_1(p,z')=z'^d$. Finally, we have
\begin{equation}
    \tilde G_{\rm ghost}(z,z',p)=z'^d\Theta(z-z').
\end{equation}
The ghost propagator vanishes as $z,z'\to0$, however, note that in this case it goes as a constant as $z\to\infty$ (for fixed position $z'$). It is also worth noting that it is independent of $p$ and asymmetric under the exchange of $z$ and $z'$.

\subsubsection{Photon propagator}

The photon propagator satisfies
\begin{equation}
    \left(g^{\mu\nu}\nabla^\sigma\nabla_\sigma-\nabla^\mu\nabla^\nu-\frac{1}{\xi}n^\nu n^\mu\right)G_{\mu\rho}(x,y)=-\frac{1}{\sqrt g}\delta^\nu_\rho\delta^{(d+1)}(x-y).
\end{equation}
In axial gauge, the equations for the photon propagator are simple enough to solve without using the BRST constraint \eqref{brst constraint}, so we will not use it here. However, we will still check that it is satisfied. We now solve for the photon propagator. In momentum space, we get the following systems of equations for the components of the photon propagator,
\begin{equation}
    \left(z^2\partial_z^2-(d-3)z\partial_z-z^2p^2\right)A=-z^{d-1}\delta(z-z'),
\end{equation}
\begin{alignat}{3}
    &z^2B''-(d-3)zB'-z^2p&&C_2'+(d-3)zpC_2&&=-z^{d-1}\delta(z-z'),\label{axial system 1 eq 2}\\
    &z^2pB'-\left(z^2p^2+\frac 1\xi\right)C_2&&=0,\label{axial system 1 eq 1}\\
    &z^2C_1''-(d-3)zC_1'-z^2p&&D'+(d-3)zpD&&=0,\label{axial system 2 eq 2}\\
    &z^2pC_1'-\left(z^2p^2+\frac 1\xi\right)D&&=-z^{d-1}\delta(z-z').\label{axial system 2 eq 1}
\end{alignat}
The equation for $A$ has already been solved in (\ref{sol for A}). The equations for $B, C_1, C_2, D$ are of the type \eqref{general system of eq w/out brst}-\eqref{general system of eq w/out brst 2}.  Starting from the first system of equations, from \eqref{axial system 1 eq 1} we find
\begin{equation}
    C_2=\frac{z^2p\xi}{1+z^2p^2\xi}B',\label{axial equation for C_2}
\end{equation}
and plugging this into (\ref{axial system 1 eq 2}) gives an equation for $B$,
\begin{equation}
    \frac{z^2}{1+z^2p^2\xi}B''-\frac{z[(d-3)+(d-1)z^2p^2\xi]}{(1+z^2p^2\xi)^2}B'=-z^{d-1}\delta(z-z').
\end{equation}
This equation can now be solved similarly to the equation for $A$. First, solve this equation for $B$ when $z\neq z'$, then impose regularity as $z\to\infty$, impose the delta-function constraints, {\it i.e.}\ continuity of $B$ and jump discontinuity of $B'$ at $z=z'$, and finally (unlike in the case of  $A$ where imposing regularity as $z\to0$ was enough to produce the correct answer) we need to impose that $B\sim z^{d-2}$ as $z\to0$ because the bulk-to-bulk propagator needs to behave as a normalizable mode close to the boundary. With these, we finally obtain that
\begin{equation}
    B=\begin{cases}
        \frac{1}{d-2}z^{d-2}+\frac{1}{d}\xi p^2z^d \,\, , \quad z<z'\\
        \frac{1}{d-2}z'^{d-2}+\frac{1}{d}\xi p^2z'^d, \quad z>z'
    \end{cases}.
\end{equation}
Now, from (\ref{axial equation for C_2}) we find
\begin{equation}
    C_2=\xi pz^{d-1}\Theta(z'-z).
\end{equation}
Moving on to the second system of equations, we can rewrite (\ref{axial system 2 eq 2}) as
\begin{equation}
    z^2(C_1'-pD)'-(d-3)z(C_1'-pD)=0,
\end{equation}
which yields
\begin{equation}
    C_1'-pD=c(z',p)z^{d-3},\label{axial equation for D}
\end{equation}
where $c(z',p)$ is an integration constant. Using this equation to express $D$ in terms of $C_1$ and plugging it into equation (\ref{axial system 2 eq 1}) gives
\begin{equation}
    C_1'-(1+z^2p^2\xi)c(z',p)z^{d-3}=p\xi z^{d-1}\delta(z-z').
\end{equation}
When $z\neq z'$ this equation can be solved for $C_1$. Imposing regularity for $C_1$ as $z\to\infty$ implies that $c(z',p)=0$ and thus (\ref{axial equation for D}) becomes $C_1'=pD$.  Jump discontinuity of $C_1$ (rather than $C_1'$ because this is now, unlike in the cases above, a first-order differential equation, not second-order) at $z=z'$ and the condition that $C_1$ goes to zero as $z\to0$ gives
\begin{equation}
    C_1=\xi pz'^{d-1}\Theta(z-z').
\end{equation}
Finally, from (\ref{axial equation for D}) we get
\begin{equation}
    D=\xi z^{d-1}\delta(z-z').
\end{equation}

Putting everything together, we have the gauge field propagator given by
\begin{equation}
    \begin{cases}
        \tilde{G}_{ij}(z,z',\vec p)&=\left(\delta_{ij}-\frac{p_ip_j}{p^2}\right)\begin{cases}
        (zz')^{\frac {d-2}2}K_{\frac{d-2}{2}}(p z')I_{\frac{d-2}{2}}(p z), \quad z<z'\\
        (zz')^{\frac {d-2}2}I_{\frac{d-2}{2}}(p z')K_{\frac{d-2}{2}}(p z), \quad z>z'
    \end{cases}\\
    &\qquad \qquad +\frac{p_ip_j}{p^2}\begin{cases}
        \frac{1}{d-2}z^{d-2}+\frac{1}{d}\xi p^2z^d \,\, , \quad z<z'\\
        \frac{1}{d-2}z'^{d-2}+\frac{1}{d}\xi p^2z'^d, \quad z>z'
    \end{cases}\\
        \tilde{G}_{i0}(z,z',\vec p)&=\frac{-i p_i}{p}\xi z'^{d-2}(pz')\Theta(z-z')\\
        \tilde{G}_{0i}(z,z',\vec p)&=\frac{i p_i}{p}\xi z^{d-2}(pz)\Theta(z'-z)\\
        \tilde{G}_{00}(z,z',\vec p)&=\xi z^{d-2}z\delta(z-z')
    \end{cases}.
\end{equation}

It is easy to see that the exchange property (\ref{exhange property mom space}) is satisfied. We are left to check that the BRST constraint (\ref{brst constraint}) is satisfied. In this case, equation (\ref{brst constraint}) becomes
\begin{equation}
    n^\mu G_{\mu\nu}(x,y)=\xi\partial_{y,\nu}G_{\rm ghost}(y,x),
\end{equation}
which in momentum space is simply
\begin{align}
    zpC_2&=\xi p^2z^d\Theta(z'-z),\\
    zpD&=\xi pz^d\delta(z-z').
\end{align}
We immediately see from the above solution that these equations are satisfied. In fact, in this case the BRST constraint (\ref{brst constraint}) simply fixes two of the components of the propagator, so we could have used it from the beginning to simplify the systems of equations that the components of the photon propagator have to satisfy.

Note that neither the ghost propagator nor the longitudinal part of the gauge field propagator vanish at large separation in this gauge. They rather limit to a constant.  We believe the origin of this is the boundary condition we use (Dirichlet) and the fact that Poincare coordinates do not cover the entire EAdS. Specifically, they do not include the boundary point at spatial infinity, which is now at $z=\infty$, and the behaviour of the propagators at large $z$ should account for the Dirichlet boundary conditions at that point. 

When $\xi=0$, which corresponds to setting $n^\mu A_\mu=zA_0=0$ so $A_0=0$ and we are in the axial gauge, the propagator is simpler and given by
\begin{equation}
    \begin{cases}
        \tilde{G}_{ij}(z,z',\vec p)&=\left(\delta_{ij}-\frac{p_ip_j}{p^2}\right)\begin{cases}
        (zz')^{\frac {d-2}2}K_{\frac{d-2}{2}}(p z')I_{\frac{d-2}{2}}(p z), \quad z<z'\\
        (zz')^{\frac {d-2}2}I_{\frac{d-2}{2}}(p z')K_{\frac{d-2}{2}}(p z), \quad z>z'
    \end{cases}+\frac{p_ip_j}{p^2}\begin{cases}
        \frac{1}{d-2}z^{d-2}, \quad z<z'\\
        \frac{1}{d-2}z'^{d-2}, \quad z>z'
    \end{cases}\label{axial gauge propagator}\\
        \tilde{G}_{i0}(z,z',\vec p)&=\tilde{G}_{0i}(z,z',\vec p)=\tilde{G}_{00}(z,z',\vec p)=0
    \end{cases}.
\end{equation}
This is in agreement (up to an overall minus sign which is just a difference in conventions) with the form computed in \cite{Marotta:2024sce} and is equivalent to the solution found in \cite{Raju:2011mp} (see appendix \ref{appendix integral representation} for the direct comparison).

\subsection{Propagator in Coulomb gauge}\label{section Coulomb gauge}

In analogy with flat space, the Coulomb (or `boundary-transverse') gauge in Poincare coordinates\footnote{Since this is a non-covariant gauge (we are using a partial derivative instead of a covariant derivative, and we single out the radial coordinate) we need to specify in which coordinates we are.} is $g^{ij}\partial_j A_i=0$ (the sum is over the boundary coordinates only). Thus, choosing $F(A)=g^{ij}\partial_j A_i$ in Poincare coordinates, the action (\ref{action}) becomes
\begin{equation}
    S=\int \mathrm{d}^{d+1}x \sqrt{g}\left[\frac14 F^{\mu\nu}F_{\mu\nu}+\frac{1}{2\xi}\left(g^{ij}\partial_j A_i\right)^2+b g^{ij}\partial_j \partial_i c\right].
\end{equation}

\subsubsection{Ghost propagator}

The ghost propagator satisfies
\begin{equation}
    g^{ij}\partial_j \partial_i G_{\rm ghost}(x,y)=\frac{1}{\sqrt g}\delta^{(d+1)}(x-y).
\end{equation}
In momentum space this becomes
\begin{equation}
    \left(-p^2z^2\right)\tilde G_{\rm ghost}(z,z',p)=z^{d+1}\delta(z-z'),
\end{equation}
so
\begin{equation}
    \tilde G_{\rm ghost}(z,z',p)=-\frac{z^{d-1}}{p^2}\delta(z-z').
\end{equation}

\subsubsection{Photon propagator}

The photon propagator satisfies
\begin{equation}
    \left(g^{\mu\nu}\nabla^\sigma\nabla_\sigma-\nabla^\mu\nabla^\nu+\frac{1}{\xi}g^{\nu j}g^{\mu i}\partial_{j}\partial_{i}\right)G_{\mu\rho}(x,y)=-\frac{1}{\sqrt g}\delta^\nu_\rho\delta^{(d+1)}(x-y)\, ,
\end{equation}
and the BRST constraint (\ref{brst constraint}) becomes
\begin{equation}
    g^{ij}\partial_j G_{i\nu}(x,y)=\xi\partial_{y,\nu}G_{\rm ghost}(y,x).
\end{equation}
In momentum space, we get the systems of equations
\begin{equation}
    \left(z^2\partial_z^2-(d-3)z\partial_z-z^2p^2\right)A=-z^{d-1}\delta(z-z'),
\end{equation}
\begin{align}
    z^2B''-(d-3)zB'-z^2pC_2'+(d-3)zpC_2+\frac{1}{\xi}\left(-z^2p^2B\right)&=-z^{d-1}\delta(z-z'),\\
    -z^2p^2C_2+z^2pB'&=0,\\
    -z^2p^2B&=\xi p^2 \tilde{G}_{\rm ghost}(z',z,p),\label{coulomb system 1 eq 3}
\end{align}
\begin{align}
    z^2C_1''-(d-3)zC_1'-z^2pD'+(d-3)zpD+\frac{1}{\xi}\left(-z^2p^2C_1\right)&=0,\\
    -z^2p^2D+z^2pC_1'&=-z^{d-1}\delta(z-z'),\\
    -z^2p^2C_1&=\xi p \partial_{z'} \tilde{G}_{\rm ghost}(z',z,p),\label{coulomb system 2 eq 3}
\end{align}
which is indeed of the type (\ref{general system of eq})-\eqref{general system of eq 2}. The equation for $A$ has already been solved in (\ref{sol for A}). As in the case of the axial gauge from section \ref{section axial gauge}, the systems of equations for the other components can be solved without using the BRST constraint, in this case equations (\ref{coulomb system 1 eq 3}) and (\ref{coulomb system 2 eq 3}), or, alternatively, we can use these equations to find $B$ and $C_1$ directly, use one of the other equations to find $C_2$ and $D$ respectively, and finally check that the remaining equation is satisfied. We proceed using the former approach in what follows. We can rewrite the above systems of equations as follows,
\begin{align}
    z^2(B'-pC_2)'-(d-3)z(B'-pC_2)+\frac{1}{\xi}\left(-z^2p^2B\right)&=-z^{d-1}\delta(z-z'),\label{coulomb system 1 eq 2}\\
    B'-pC_2&=0,\label{coulomb system 1 eq 1}\\
    B&=\xi \frac{z^{d-3}}{p^2}\delta(z-z'),\label{coulomb system 1 eq 3 again}
\end{align}
\begin{align}
    z^2(C_1'-pD)'-(d-3)z(C_1'-pD)+\frac{1}{\xi}\left(-z^2p^2C_1\right)&=0,\label{coulomb system 2 eq 2}\\
    C_1'-pD&=-\frac{z^{d-3}}{p}\delta(z-z'),\label{coulomb system 2 eq 1}\\
    C_1&=\xi \frac{z^{d-3}}{p^3}\partial_{z'}\delta(z'-z).\label{coulomb system 2 eq 3 again}
\end{align}
Equations (\ref{coulomb system 1 eq 3 again}) and (\ref{coulomb system 2 eq 3 again}) are just the BRST constraints. We will solve for the remaining components using only the first two equations from each system and show that the results agree with (\ref{coulomb system 1 eq 3 again}) and (\ref{coulomb system 2 eq 3 again}). Starting from the first system, using (\ref{coulomb system 1 eq 1}) in (\ref{coulomb system 1 eq 2}) we find
\begin{equation}
    B=\xi \frac{z^{d-3}}{p^2}\delta(z-z'),
\end{equation}
which is the same as (\ref{coulomb system 1 eq 3 again}). Using (\ref{coulomb system 1 eq 1}) we find
\begin{equation}
    C_2=\xi \frac{z'^{d-3}}{p^3}\partial_z\delta(z-z').
\end{equation}
Moving on to the second system, using (\ref{coulomb system 2 eq 1}) in (\ref{coulomb system 2 eq 2}) we have
\begin{equation}
    C_1=\xi \frac{z^{d-3}}{p^3}\partial_{z'}\delta(z'-z),
\end{equation}
which is again the same as (\ref{coulomb system 2 eq 3 again}). Finally, using (\ref{coulomb system 2 eq 1}) we have
\begin{equation}
    D=\frac{z^{d-3}}{p^2}\delta(z-z')+\xi\frac{1}{p^2}\partial_z\partial_{z'}\left(\frac{z^{d-3}}{p^2}\delta(z-z')\right).
\end{equation}
The solution is now complete, and it can be checked that it also satisfies the exchange property (\ref{exhange property mom space}).

The full solution is
\begin{equation}
    \begin{cases}
        \tilde{G}_{ij}(z,z',\vec p)&=\left(\delta_{ij}-\frac{p_ip_j}{p^2}\right)\begin{cases}
        (zz')^{\frac {d-2}2}K_{\frac{d-2}{2}}(p z')I_{\frac{d-2}{2}}(p z), \quad z<z'\\
        (zz')^{\frac {d-2}2}I_{\frac{d-2}{2}}(p z')K_{\frac{d-2}{2}}(p z), \quad z>z'
    \end{cases}+\frac{p_ip_j}{p^2}\xi \frac{z^{d-3}}{p^2}\delta(z-z')\\
        \tilde{G}_{i0}(z,z',\vec p)&=\frac{-i p_i}{p}\xi \frac{z^{d-3}}{p^3}\partial_{z'}\delta(z'-z)\\
        \tilde{G}_{0i}(z,z',\vec p)&=\frac{i p_i}{p}\xi \frac{z'^{d-3}}{p^3}\partial_z\delta(z-z')\\
        \tilde{G}_{00}(z,z',\vec p)&=\frac{z^{d-3}}{p^2}\delta(z-z')+\xi\frac{1}{p^2}\partial_z\partial_{z'}\left(\frac{z^{d-3}}{p^2}\delta(z-z')\right)
    \end{cases}.
\end{equation}
For $\xi=0$, when we are properly in the Coulomb gauge, this becomes
\begin{equation}
    \begin{cases}
        \tilde{G}_{ij}(z,z',\vec p)&=\left(\delta_{ij}-\frac{p_ip_j}{p^2}\right)\begin{cases}
        (zz')^{\frac {d-2}2}K_{\frac{d-2}{2}}(p z')I_{\frac{d-2}{2}}(p z), \quad z<z'\\
        (zz')^{\frac {d-2}2}I_{\frac{d-2}{2}}(p z')K_{\frac{d-2}{2}}(p z), \quad z>z'
    \end{cases}\label{coulomb gauge propagator}\\
        \tilde{G}_{i0}(z,z',\vec p)&=\tilde{G}_{0i}(z,z',\vec p)=0\\
        \tilde{G}_{00}(z,z',\vec p)&=\frac{z^{d-3}}{p^2}\delta(z-z')
    \end{cases}.
\end{equation}
This propagator agrees with the propagator in \cite{Liu:1998ty}. The propagator is very similar to the axial gauge result (\ref{axial gauge propagator}), with the difference here being that $\tilde{G}_{ij}(z,z',\vec p)$ does not contain a longitudinal part but $\tilde{G}_{00}(z,z',\vec p)$ is non-zero. The simplicity of the propagator in these gauges is due to the fact that, in Poincare coordinates, these gauge choices nicely exploit the geometry of AdS. This will not be the case for the propagator in the general covariant gauge which we compute in section \ref{section standard covariant gauge}.

\subsection{Relation between Coulomb and axial gauge}\label{sec: axial_to_Coulomb}

We have seen by explicit computation that the propagators in axial and Coulomb gauge are closely related to each other, and we would like to explain this relation in this subsection. First, we note that both gauges have a residual gauge invariance, $\delta A_\mu = \partial_\mu \lambda$, that preserves the gauge,
\begin{align}
    {\rm Axial}: & \qquad    \delta A_0=0 \qquad \Rightarrow\quad \partial_z \lambda_{\rm axial}(z,\vec x) =0 \qquad \Rightarrow\quad \lambda_{\rm axial} = \lambda_{\rm axial}(\vec{x}), \nonumber \\
    {\rm Coulomb}: &  \quad   g^{ij} \partial_j \delta A_i=0 \,\,\,\,\,\, \Rightarrow\quad g^{ij} \partial_j \partial_i \lambda_{\rm C} (z, \vec{x}) =0\, .
\end{align}
These residual transformations are not sufficient to impose both gauges at the same time. However, on-shell one can solve the linearized field equations in the Lorenz gauge (also called the Landau gauge), $\nabla^\mu A_\mu=0$, and then show that there is enough residual gauge invariance to set $A_z=0$ \cite{Marotta:2024sce}. After this is done, the Lorenz condition reduces to the Coulomb condition. Such a choice has appeared in previous computations: in the axial gauge computation in \cite{Raju:2011mp} the solutions were taken to be transverse, and in the Coulomb gauge computation in \cite{Liu:1998ty} the gauge field was taken to satisfy $A_z=0$.
 
We will now explicitly relate the propagators in the two gauges. We will focus on the $\xi=0$ case on both sides. Additional gauge transformations can generate the $\xi$-dependence. In axial gauge, $A_z=0$, the propagator is not transverse and we would like to make a gauge transformation to achieve this,
\begin{equation} \label{axial_to_Coulomb}
    \hat{A}_\mu = A^{\rm axial}_\mu + \partial_\mu \lambda.
\end{equation}
As this affects only the longitudinal part of the gauge field, we will focus on the longitudinal part of the correlators in the following and we will indicate this by $\langle \ \rangle_L$. We view $\lambda$ as a quantum operator and would like to determine the correlators of $\lambda$ with $A_i^{\rm axial}$ and itself such that $\hat{A}_i$ is transverse, {\it i.e.}
\begin{equation}
    \left\langle\hat{A}_i(z,\vec x)\hat{A}_j(z',\vec y)\right\rangle_L=\left \langle \hat{A}_i(z, \vec{x}) A_j^{\rm axial} (z', \vec{y})\right \rangle_L+\partial_{y_j}\left \langle \hat{A}_i(z, \vec{x}) \lambda(z', \vec{y})\right \rangle_L=0.
\end{equation}
We will now show that there exist correlators of $\lambda$ and $A_i^{\rm axial}$ such that both $\langle \hat{A}_i(z, \vec{x}) A_j^{\rm axial} (z', \vec{y}) \rangle_L$ and $\langle \hat{A}_i(z, \vec{x}) \lambda(z', \vec{y}) \rangle_L$ vanish. In momentum space, we have that
\begin{align}
    &\left \langle \hat{A}_i(z, \vec{p}) A_j^{\rm axial} (z', -\vec{p})\right \rangle_L = 0 \quad \Leftrightarrow \quad \left \langle A^{\rm axial}_i(z, \vec{p}) A_j^{\rm axial} (z',- \vec{p})\right \rangle_L + (-ip_i) \left \langle \lambda(z, \vec{p}) A_j^{\rm axial} (z',- \vec{p})\right \rangle_L =0 \nonumber \\
    &\Leftrightarrow \left \langle \lambda(z, \vec{p}) A_j^{\rm axial} (z', -\vec{p})\right \rangle_L = -i\frac{p_j}{p^2}\frac{1}{d-2}\left(z'^{d-2}\Theta(z-z')+z^{d-2}\Theta(z'-z)\right),
\end{align}
and 
\begin{align}
    &\left \langle \hat{A}_i(z, \vec{p}) \lambda(z', -\vec{p})\right \rangle_L = 0 \quad \Leftrightarrow \quad \left \langle A^{\rm axial}_i(z, \vec{p}) \lambda (z', -\vec{p})\right \rangle_L + (-ip_i) \left \langle \lambda(z, \vec{p}) \lambda (z', -\vec{p})\right \rangle =0 \nonumber \\
    &\Leftrightarrow \left \langle \lambda(z, \vec{p}) \lambda (z', -\vec{p})\right \rangle= \frac{1}{d-2}\frac{1}{p^2}\left(z'^{d-2}\Theta(z-z')+z^{d-2}\Theta(z'-z)\right).
\end{align}
These further imply that
\begin{equation}
    \left \langle \hat{A}_0(z, \vec{p}) \hat{A}_0(z', -\vec{p})\right \rangle =\partial_z \partial_{z'} \left \langle \lambda(z, \vec{p}) \lambda (z', -\vec{p})\right \rangle = \frac{z^{d-3}}{p^2}\delta(z-z'),
\end{equation}
which is exactly the $00$-component of the Coulomb gauge propagator. Additionally, one can check that
\begin{equation}
    \left \langle \hat{A}_0(z, \vec{p}) \hat{A}_j(z', -\vec{p})\right \rangle=\left \langle \hat{A}_i(z, \vec{p}) \hat{A}_0(z', -\vec{p})\right \rangle=0.
\end{equation}
In other words, the gauge transformation \eqref{axial_to_Coulomb} maps the axial gauge propagator to the Coulomb gauge propagator.

\subsection{Propagator in covariant gauge}\label{section standard covariant gauge}

We now consider the standard covariant gauge fixing term. Choosing $F(A)=\nabla^\mu A_\mu$, the action (\ref{action}) becomes
\begin{equation}
    S=\int \mathrm{d}^{d+1}x \sqrt{g}\left[\frac14 F^{\mu\nu}F_{\mu\nu}+\frac{1}{2\xi}\left(\nabla^\mu A_\mu\right)^2+b\nabla^\mu\partial_\mu c\right].
\end{equation}

\subsubsection{Ghost propagator}

The ghost propagator satisfies
\begin{equation}
    \nabla^\mu\partial_\mu G_{\rm ghost}(x,y)=\frac{1}{\sqrt g}\delta^{(d+1)}(x-y).
\end{equation}
In momentum space this becomes
\begin{equation}
    \left(z^2\partial_z^2-(d-1)z\partial_z-p^2z^2\right)\tilde G_{\rm ghost}(z,z',p)=z^{d+1}\delta(z-z').
\end{equation}
This equation can be solved in exactly the same way as equation (\ref{eq for A}). The solution in this case is
\begin{equation}
    \tilde G_{\rm ghost}(z,z',p)=-\left((zz')^{\frac d2}I_{\frac{d}{2}}(p z')K_{\frac{d}{2}}(p z)\right)\Theta(z-z')-\left((zz')^{\frac d2}K_{\frac{d}{2}}(p z')I_{\frac{d}{2}}(p z)\right)\Theta(z'-z).
\end{equation}

\subsubsection{Photon propagator}

The photon propagator satisfies
\begin{equation}
    \left(g^{\mu\nu}\nabla^\sigma\nabla_\sigma-\nabla^\mu\nabla^\nu+\frac{1}{\xi}\nabla^\nu\nabla^\mu\right)G_{\mu\rho}(x,y)=-\frac{1}{\sqrt g}\delta^\nu_\rho\delta^{(d+1)}(x-y)
\end{equation}
and the BRST constraint (\ref{brst constraint}) becomes
\begin{equation}
    \nabla^\mu G_{\mu\nu}(x,y)=\xi\partial_{y,\nu}G_{\rm ghost}(y,x).
\end{equation}
In momentum space, we get the systems of equations
\begin{equation}
    \left(z^2\partial_z^2-(d-3)z\partial_z-z^2p^2\right)A=-z^{d-1}\delta(z-z'),
\end{equation}
\begin{align}
    z^2B''-(d-3)zB'-z^2pC_2'+(d-3)zpC_2\quad\quad\quad\quad\quad\\
    +\frac{1}{\xi}\left(z^2pC_2'-(d-1)zpC_2-z^2p^2B\right)&=-z^{d-1}\delta(z-z'),\nonumber\\
    -z^2p^2C_2+z^2pB'\quad\quad\quad\quad\quad\quad\quad\quad\quad\quad\quad\quad\quad\quad\quad\quad\\
    +\frac{1}{\xi}\left(z^2C_2''-(d-3)zC_2'-(d-1)C_2-z^2pB'-2zpB\right)&=0,\nonumber\\
    z^2pC_2'-(d-1)zpC_2-z^2p^2B&=\xi p^2 \tilde{G}_{\rm ghost}(z',z,p),
\end{align}
\begin{align}
    z^2C_1''-(d-3)zC_1'-z^2pD'+(d-3)zpD\quad\quad\quad\quad\quad\\
    +\frac{1}{\xi}\left(z^2pD'-(d-1)zpD-z^2p^2C_1\right)&=0,\nonumber\\
    -z^2p^2D+z^2pC_1'\quad\quad\quad\quad\quad\quad\quad\quad\quad\quad\quad\quad\quad\quad\quad\quad\\
    +\frac{1}{\xi}\left(z^2D''-(d-3)zD'-(d-1)D-z^2pC_1'-2zpC_1\right)&=-z^{d-1}\delta(z-z'),\nonumber\\
    z^2pD'-(d-1)zpD-z^2p^2C_1&=\xi p \partial_{z'} \tilde{G}_{\rm ghost}(z',z,p),
\end{align}
which is indeed of the type (\ref{general system of eq})-\eqref{general system of eq 2}. The equation for $A$ has already been solved in (\ref{sol for A}).

Using substitution and manipulating the equations we find
\begin{align}
    z^2B''-(d-1)zB'-z^2p^2B&=-z^{d-1}\delta(z-z')\\
    &+(\xi-1)p^2\tilde{G}_{\rm ghost}(z',z,p)+\frac{2}{z}\partial_z\tilde{G}_{\rm ghost}(z',z,p),\nonumber\\
    C_2&=\frac{1}{p}B'+\frac{1}{z^2p}\partial_z\tilde{G}_{\rm ghost}(z',z,p),
\end{align}
\begin{align}
    z^2D''-(d-1)zD'+(d-1-z^2p^2)D&=-z^{d-1}\delta(z-z')\\
    &-\frac{2\xi}{z}\partial_{z'}\tilde{G}_{\rm ghost}(z',z,p)+(\xi-1)\partial_z\partial_{z'}\tilde{G}_{\rm ghost}(z',z,p),\nonumber\\
    C_1&=\frac{1}{p}D'-\frac{d-1}{zp}D-\frac{\xi}{z^2p}\partial_{z'}\tilde{G}_{\rm ghost}(z',z,p).
\end{align}
The equations for $B$ and $D$ can now be solved similarly as the equation for $A$ was solved in (\ref{sol for A}),\footnote{The only difference here is that the equations at separated points, $z\neq z'$, are now inhomogeneous equations. One needs to obtain and write down both the complementary function and the particular integral for each branch ($z>z'$ and $z<z'$), and only then impose the boundary conditions (regularity at infinity and normalizable fall off as $z,z'\to0$) and the delta-function constraint.} and then these can be used to find $C_2$ and $C_1$ (which we check that they satisfy the right boundary conditions). It then needs to be checked that the final solutions satisfy all the equations and the exchange property (\ref{exhange property mom space}).

\subsubsection*{Comments}

The homogeneous solutions of the equations for $B$ and $D$ are expressed in terms of Bessel functions similar to the solution for $\tilde{G}_{\rm ghost}(z,z',p)$. For $d$ odd, these are Bessel functions of half-integer order, so they simplify to simple functions. For $d$ even the order of the Bessel functions is an integer, so their form does not simplify. In what follows we show the final results for $d=3$ and we include a Mathematica file which contains the derivations and solutions for specific values of $d$. For odd $d$ the particular integrals of the differential equations satisfied by $B$ and $D$ are simple enough that Mathematica can handle. For even $d$ these become integrals over products of Bessel functions of integer order which Mathematica can not process. In this case we rewrite the products of Bessel functions as Meijer G-functions which can then be integrated. This leads to very very complicated expressions for the propagator for even $d$ which, in its current form, Mathematica can not simplify further. On a different note, we would like to mention here that, as also observed in \cite{Glavan:2022nrd} and in \cite{Ciccone:2024guw}, the propagator is simplest in the Fried-Yennie gauge $\xi=d/(d-2)$ \cite{PhysRev.112.1391,Yennie:1961ad}.

\subsubsection{Solution in $d=3$}

We list here the expressions for $B,C_2,C_1,$ and $D$. We use the following notation in what follows,
\begin{equation}
    {\cal A}={\cal A}_{(1)}\Theta(z-z')+{\cal A}_{(2)}\Theta(z'-z),
\end{equation}
where ${\cal A}=\{B,C_2,C_1,D\}$. One can see immediately that for $\xi=3$, {\it i.e.}\ in the Fried-Yennie gauge, all expressions simplify dramatically. We have
\begingroup
\allowdisplaybreaks
\begin{align}
    B_{(1)}(z,z')&=\begin{aligned}[t]
    &\frac{e^{-p (z-z')} \left(p (z-z')+2\right)-e^{-p (z+z')}\left(p (z+z')+2\right)}{2 p}\\
    &-(\xi-3)\frac{1}{6p}\Bigg[(pz+1)(pz'-1)e^{-p(z-z')}\text{Ei}(-2 p z')\\
    &\quad\quad\quad\quad\quad\,\,\,+(pz+1)(pz'+1)e^{-p(z+z')}\text{Ei}(2 p z')\\
    &\quad\quad\quad\quad\quad\,\,\,+(pz-1)\left((pz'+1)e^{-p(z'-z)}+(pz'-1)e^{p(z+z')}\right)\text{Ei}(-2 p z)\Bigg]\, ;
    \end{aligned}\\
    B_{(2)}(z,z')&=B_{(1)}(z',z)\, ;\\
    C_{2(1)}(z,z')&=\begin{aligned}[t]
    &\frac{e^{-p(z-z')} \left(p^2 z (z'-z)+p (z'-z)-1\right)+e^{-p (z+z')} \left(p^2 z (z+z')+p (z+z')+1\right)}{2 p^2 z}\\
    &-(\xi-3)\frac{1}{6p^2z}\Bigg[\left(p^2z^2+p^2z(z-z')+p(z-z')+1\right)e^{-p(z-z')}\\
    &\quad\quad\quad\quad\quad\quad\,\,-\left(p^2z^2+p^2z(z+z')+p(z+z')+1\right)e^{-p(z+z')}\\
    &\quad\quad\quad\quad\quad\quad\,\,-p^2z^2(pz'-1)e^{-p(z-z')}\text{Ei}(-2 p z')\\
    &\quad\quad\quad\quad\quad\quad\,\,-p^2z^2(pz'+1)e^{-p(z+z')}\text{Ei}(2 p z')\\
    &\quad\quad\quad\quad\quad\quad\,\,+p^2z^2\left((pz'+1)e^{-p(z'-z)}+(pz'-1)e^{p(z+z')}\right)\text{Ei}(-2 p z)\Bigg]\, ;
    \end{aligned}\\
    C_{2(2)}(z,z')&=\begin{aligned}[t]
    &\frac{e^{-p(z'-z)} \left(p^2 z (z'-z)+p (z-z')-1\right)+e^{-p (z+z')} \left(p^2 z (z+z')+p (z+z')+1\right)}{2 p^2 z}\\
    &-(\xi-3)\frac{1}{6p^2z}\Bigg[\left(p^2z^2+p^2z(z-z')-p(z-z')+1\right)e^{-p(z'-z)}\\
    &\quad\quad\quad\quad\quad\quad\,\,-\left(p^2z^2+p^2z(z+z')+p(z+z')+1\right)e^{-p(z+z')}\\
    &\quad\quad\quad\quad\quad\quad\,\,+p^2z^2(pz'+1)e^{-p(z'-z)}\text{Ei}(-2 p z)\\
    &\quad\quad\quad\quad\quad\quad\,\,-p^2z^2(pz'+1)e^{-p(z+z')}\text{Ei}(2 p z)\\
    &\quad\quad\quad\quad\quad\quad\,\,-p^2z^2(pz'-1)\left(e^{-p(z-z')}-e^{p(z+z')}\right)\text{Ei}(-2 p z')\Bigg]\, ;
    \end{aligned}\\
    C_{1(1)}(z,z')&=C_{2(2)}(z',z)\, ;\\
    C_{1(2)}(z,z')&=C_{2(1)}(z',z)\, ;\\
    D_{(1)}(z,z')&=\begin{aligned}[t]
    &\frac{e^{-p (z-z')} (z^2+z'^2-pzz'(z-z'))-e^{-p (z+z')}\left(z^2+z'^2+p zz'(z+z')\right)}{2 p z z'}\\
    &-(\xi-3)\frac{1}{6pzz'}\Bigg[\left(z^2(pz'-1)-z'^2(pz+1)\right)e^{-p(z-z')}\\
    &\quad\quad\quad\quad\quad\quad\,\,\,\,+\left(z^2(pz'+1)+z'^2(pz+1)\right)e^{-p(z+z')}\\
    &\quad\quad\quad\quad\quad\quad\,\,\,\,-p^2z^2z'^2e^{-p(z-z')}\text{Ei}(-2 p z')\\
    &\quad\quad\quad\quad\quad\quad\,\,\,\,+p^2z^2z'^2e^{-p(z+z')}\text{Ei}(2 p z')\\
    &\quad\quad\quad\quad\quad\quad\,\,\,\,-p^2z^2z'^2\left(e^{-p(z'-z)}-e^{p(z+z')}\right)\text{Ei}(-2 p z)\Bigg]\, ;
    \end{aligned}\\
    D_{(2)}(z,z')&=D_{(1)}(z',z)\, ;
\end{align}
\endgroup
where $\text{Ei}(x)$ is the exponential function.\footnote{The exponential function for real non-zero values $x$ is defined by
\begin{equation}
    \text{Ei}(x) = \int_{-x}^\infty \frac{e^{-t}}{t} dt\, . 
\end{equation}
}

\subsubsection{Flat space limit}

Recall that the Euclidean flat space propagator in momentum space is given by (\ref{covariant flat space prop}),
\begin{equation}
    \tilde G_{\mu\rho}(p)=\frac{1}{p^2}\left(\delta_{\mu\rho}-\left(1-\xi\right)\frac{p_\mu p_\rho}{p^2}\right).
\end{equation}
This can be written in `semi-momentum' space by Fourier transforming back one of the coordinates,
\begin{equation}
    \bar G_{\mu\nu}(\tau,\tau',\vec p)=\int\frac{\mathrm{d}p_0}{2\pi}\tilde{G}_{\mu\nu}(p)e^{-i p_0\cdot(\tau-\tau')},\label{semi momentum FT}
\end{equation}
where $p=(p_0,\vec p)$. Similarly to how we decomposed the propagator in EAdS in equation (\ref{ads decomposition}), we decompose the flat space propagator as
\begin{equation}
    \begin{cases}
        \bar{G}_{ij}(\tau,\tau',\vec p)&=A^{\rm flat}\left(\delta_{ij}-\frac{p_ip_j}{p^2}\right)+B^{\rm flat}\frac{p_ip_j}{p^2}\\
        \bar{G}_{i0}(\tau,\tau',\vec p)&=C_1^{\rm flat}\frac{-ip_i}{p}, \quad \bar{G}_{0i}(\tau,\tau',\vec p)=C_2^{\rm flat}\frac{ip_i}{p}\\
        \bar{G}_{00}(\tau,\tau',\vec p)&=D^{\rm flat}
    \end{cases},
\end{equation}
where now the form factors are functions of $\tau,\tau',$ and $|\vec p|$. We will write $|\vec p|\equiv p$ from now on for the rest of this section. Completing the Fourier transform in (\ref{semi momentum FT}), we get
\begin{align}
    A^{\rm flat}(\tau,\tau')&=\frac{1}{2p}\left(e^{-p(\tau-\tau')}\Theta(\tau-\tau')+(\tau\leftrightarrow\tau')\right),\\
    B^{\rm flat}(\tau,\tau')&=\frac{1}{4p}\left(e^{-p(\tau-\tau')}(1-p(\tau-\tau'))\Theta(\tau-\tau')+(\tau\leftrightarrow\tau')\right)\nonumber\\
    &+\frac{\xi}{4p}\left(e^{-p(\tau-\tau')}(1+p(\tau-\tau'))\Theta(\tau-\tau')+(\tau\leftrightarrow\tau')\right),\\
    C_1^{\rm flat}(\tau,\tau')&=\frac{\xi-1}{4p}\left(e^{-p(\tau-\tau')}p(\tau-\tau')-(\tau\leftrightarrow\tau')\right),\\
    C_2^{\rm flat}(\tau,\tau')&=\frac{1-\xi}{4p}\left(e^{-p(\tau-\tau')}p(\tau-\tau')-(\tau\leftrightarrow\tau')\right),\\
    D^{\rm flat}(\tau,\tau')&=\frac{1}{4p}\left(e^{-p(\tau-\tau')}(1+p(\tau-\tau'))\Theta(\tau-\tau')+(\tau\leftrightarrow\tau')\right)\nonumber\\
    &+\frac{\xi}{4p}\left(e^{-p(\tau-\tau')}(1-p(\tau-\tau'))\Theta(\tau-\tau')+(\tau\leftrightarrow\tau')\right).
\end{align}

The prescription for how to take the flat space limit for momentum space EAdS has been described in detail in \cite{Marotta:2024sce}, so we just summarize the procedure for the bulk-to-bulk propagator here. The first step is to reintroduce the AdS radius, $l_{\rm AdS}$, which has so far been set to $1$. This corresponds for the propagator to including an overall multiplicative factor of $l_{\rm AdS}^{-(d-3)}$. Next we change the $z$-coordinate to $z=l_{\rm AdS}\,e^{\tau/l_{\rm AdS}}$, where $\tau$ will be identified with the Euclidean time in the flat space limit. The flat space limit is now obtained by sending the radius of AdS to infinity, {\it i.e.}\ $l_{\rm AdS}\to\infty$. We confirm that for the solution obtained above in $d=3$ we indeed have
\begin{equation}
    \lim_{l_{\rm AdS}\to\infty}{\cal A}(l_{\rm AdS}\,e^{\tau/l_{\rm AdS}},l_{\rm AdS}\,e^{\tau'/l_{\rm AdS}})={\cal A}^{\rm flat}(\tau,\tau'),
\end{equation}
where ${\cal A}=\{A,B,C_1,C_2,D\}$. Unlike in the case considered in \cite{Marotta:2024sce} where the flat space limit of the propagator in axial gauge was studied, in this case there are no subtleties when taking the limit. In axial gauge there is a divergent piece when taking the limit $l_{\rm AdS}\to\infty$ of the propagator which matches the well-known singularity of the flat space propagator in axial gauge.
\section{What is special about the Fried-Yennie gauge?}\label{section position space}

In this section, we discuss what is special about the Fried-Yennie gauge, {\it i.e.}\ when the corresponding gauge fixing term is $\frac{1}{2\xi}\left(\nabla^\mu A_\mu\right)^2$ with $\xi=\frac{d}{d-2}$. The Fried-Yennie gauge was originally introduced in flat space\cite{PhysRev.112.1391} and it was shown in \cite{Yennie:1961ad} that perturbation theory in QED exhibits improved IR behaviour in this gauge. Discussions of this gauge and explicit computations can be found, for example, in \cite{Tomozawa:1979xn,Adkins:1992zza,Adkins:1993qm,Adkins:1994uy,Eides:2001dw,Mizher:2024zag} (see also problem 2 in page 521 of \cite{Brown:1992db}.) Here we will provide additional arguments about the improved IR behavior of this gauge.

Recall that in Euclidean flat space, the photon propagator in the standard covariant gauge takes the form (\ref{covariant flat space prop}),
\begin{equation}
    \tilde G_{\mu\rho}(p)=\frac{1}{p^2}\left(\delta_{\mu\rho}-\left(1-\xi\right)\frac{p_\mu p_\rho}{p^2}\right).\label{covariant flat space prop main}
\end{equation}
For $\xi=\frac{d}{d-2}$ this becomes
\begin{equation}
    \tilde G_{\mu\rho}(p)=\frac{1}{p^2}\left(\delta_{\mu\rho}+\frac{2}{d-2}\frac{p_\mu p_\rho}{p^2}\right).
\end{equation}
There does not appear to be anything special about this gauge choice in momentum space. Usually, the standard choices are the Feynman gauge, $\xi=1$, which yields a propagator proportional to the metric, and the Landau gauge, $\xi=0$, for which the propagator is transverse to the photon momentum. The reason why the Fried-Yennie gauge is special is more transparent in position space. The position space photon propagator is given by (see for example \cite{Narain:2014oja}),
\begin{equation}
    G_{\mu\nu'}(\sigma)=\frac{\Gamma\left(\frac{d+1}{2}\right)}{(4\pi)^\frac{d+1}{2}}\left(\frac{2}{\sigma}\right)^{\frac{d-1}{2}}\left[\left(\frac{g_{\mu\nu'}}{d-1}-\frac{\nabla_\mu\sigma\nabla_{\nu'}\sigma}{2\sigma}\right)+\xi\left(\frac{g_{\mu\nu'}}{d-1}+\frac{\nabla_\mu\sigma\nabla_{\nu'}\sigma}{2\sigma}\right)\right],
\end{equation}
where $\sigma$, the Synge's world function, is given by $\sigma(x,x')=\mu(x,x')^2/2$, with $\mu(x,x')=\sqrt{(x-x')^2}$ the geodesic distance between $x$ and $x'$. Here $g_{\mu\nu'}$ is a bitensor (not to be confused with the metric tensor) known as the parallel propagator along the geodesic from $x$ to $x'$. It is however the case that for Euclidean flat space in Cartesian coordinates $g_{\mu\nu'}=\rm diag (1,\ldots,1)$.

We now change coordinates and work directly with $\mu$ instead of $\sigma$. Using the fact that in Euclidean flat space $g_{\mu\nu'}=-\nabla_\mu\nabla_{\nu'}\sigma$, we can rewrite the propagator as follows
\begin{equation}
    G_{\mu\nu'}(\mu)=\frac{\Gamma\left(\frac{d+1}{2}\right)}{4\pi^\frac{d+1}{2}}\frac{1}{\mu^{d-1}}\left[-\frac{1+\xi}{d-1}\mu\nabla_\mu\nabla_{\nu'}\mu+\frac{d-2}{d-1}\left(\xi-\frac{d}{d-2}\right)\nabla_\mu\mu\nabla_{\nu'}\mu\right].\label{flat space gauge prop in mu variable}
\end{equation}
In these variables the propagator in the Fried-Yennie gauge $\xi=\frac{d}{d-2}$ has a similarity with the momentum space Feynman propagator in that only one tensor structure has a non-zero coefficient. We may further introduce a new variable $\mu_\mu=(x-x')_\mu$, such that $\mu^2=\mu_\mu \mu^\mu$, and rewrite the above expression as
\begin{equation}
    G_{\mu\nu'}(\mu)=\frac{\Gamma\left(\frac{d+1}{2}\right)}{4\pi^\frac{d+1}{2}}\frac{1}{\mu^{d-1}}\left[\frac{1+\xi}{d-1}\left(\delta_{\mu\nu'}-\frac{\mu_\mu\mu_{\nu'}}{\mu^2}\right)+\frac{d-2}{d-1}\left(\frac{d}{d-2}-\xi\right)\frac{\mu_\mu\mu_{\nu'}}{\mu^2}\right].
\end{equation}
Contracting the above with $\mu^\mu$, we get
\begin{equation}
    \mu^\mu G_{\mu\nu'}(\mu)=\frac{\Gamma\left(\frac{d+1}{2}\right)}{4\pi^\frac{d+1}{2}}\frac{d-2}{d-1}\left(\frac{d}{d-2}-\xi\right)\frac{1}{\mu^{d-1}}\mu_{\nu'}=0 \quad \text{for} \quad \xi=\frac{d}{d-2}.\label{flat space transversality}
\end{equation}
This suggests that the propagator in the Fried-Yennie gauge is akin to the Landau gauge momentum space propagator. It is well-known that in momentum space the Landau gauge has improved UV behaviour. This is because the propagator in Landau gauge is transverse and thus it vanishes when contracted with a momentum, instead of giving a term proportional to the momentum (which can lead to divergences as $p \to \infty$) as in Feynman gauge. While in Feynman gauge such terms would cancel among different diagrams (as required by gauge invariance), in Landau gauge they vanish diagram by diagram, and one may explicitly check using power-counting that Feynman integrals in Landau gauge indeed exhibit an improved UV behaviour. In the Fried-Yennie gauge we have an exactly analogous situation but now in the IR. In a generic gauge, the gauge propagator contracted by $\mu^\mu$ will be proportional to $\mu_{\nu'}$ and such factors can lead to divergences in the IR limit, $\mu \to \infty$. Such terms are gauge dependent and they would cancel between different diagrams. In contrast, in the Fried-Yennie gauge, such large IR contributions to individual diagrams never arise because the propagator is $\mu$-transverse.

We also note here that if we rewrite $\xi$ as $\xi=\frac{d-\kappa}{d-2+\kappa}$, where now $\kappa$ can be viewed as the gauge fixing parameter, the propagator takes the form
\begin{equation}
    G_{\mu\nu'}(\mu)=\frac{\Gamma\left(\frac{d+1}{2}\right)}{4\pi^\frac{d+1}{2}}\frac{2}{d+\kappa-2}\frac{1}{\mu^{d-1}}\left(\delta_{\mu\nu'}-(1-\kappa)\frac{\mu_\mu\mu_{\nu'}}{\mu^2}\right),
\end{equation}
which looks more like (\ref{covariant flat space prop main}). With this parametrization, the Fried-Yennie gauge corresponds to $\kappa=0$. The propagator is rather simple also for $\kappa=1$,
\begin{equation}
    G_{\mu\nu'}(\mu)=\frac{\Gamma\left(\frac{d+1}{2}\right)}{2(d-1)\pi^\frac{d+1}{2}}\frac{1}{\mu^{d-1}}\delta_{\mu\nu'}.
\end{equation}
This is not surprising since $\kappa=1$ corresponds to $\xi=1$, so this is the Feynman gauge. Another, potentially interesting, choice is $\kappa=-1$. With this choice, the propagator takes the form
\begin{equation}
    G_{\mu\nu'}(\mu)=\frac{\Gamma\left(\frac{d+1}{2}\right)}{2(d-3)\pi^\frac{d+1}{2}}\frac{1}{\mu^{d-1}}\left(\delta_{\mu\nu'}-2\frac{\mu_\mu\mu_{\nu'}}{\mu^2}\right)\equiv\frac{\Gamma\left(\frac{d+1}{2}\right)}{2(d-3)\pi^\frac{d+1}{2}}\frac{I_{\mu\nu'}(x-x')}{\mu^{d-1}},
\end{equation}
where $I_{\mu\nu'}(x-x')$ is the inversion tensor. It turns out that in this gauge the $2$-point function has the form of a CFT vector primary with dimension $\Delta=\frac{d-1}{2}$. This is below the unitarity bound of a unitary CFT vector primary. Nevertheless, this gauge choice has been useful: it was used in \cite{El-Showk:2011xbs} to show that the free Maxwell theory in more than four dimensions (which is scale, but not conformally invariant) can be embedded into a non-unitary CFT.
\section{Position space propagator in covariant gauge}\label{section position space 2}

The form of the position space photon propagator in the Fried-Yennie gauge described in the previous section (in flat space) suggests that choosing the right variables would make the simplicity of the propagator in this gauge in EAdS also manifest. Assuming one can quantize the theory while maintaining AdS invariance, the propagators will depend on the (unique) AdS-invariant distance only. This is not an innocuous assumption, as the debate regarding massless scalars in dS demonstrates (see, for example, \cite{Allen:1985ux} (and \cite{Kirsten:1993ug} for a different perspective) and \cite{Glavan:2022nrd} for a corresponding discussion in the case of gauge fields). In our case, and for Dirichlet boundary conditions, we will explicitly check that the solutions we obtain (assuming the propagator depends only on the AdS invariant distance) satisfy all required consistency checks. The case of other boundary conditions, such as Neumann, is more subtle and will be addressed elsewhere.

The defining equations for the propagators are distributional and it is important to properly take this fact into account.  In momentum space the problem reduces to a one-dimensional problem which can be solved using familiar methods (as described earlier). In position space the proper way to process distributional equations is to multiply them with suitable test functions and integrate. This approach will be discussed in detail in \cite{BIStoappear}. An alternative approach is to note that the delta-function has a support at coincident points and this is a short-distance effect. As such, the behaviour of the propagator close to the coincident limit should be that of the flat space propagator. We will thus start by solving the equations at non-coincident points, and then ensure that the delta-function constraints are satisfied by fixing the constants of integration such that the propagators approach the flat space expression at the coincident limit. 

There is a unique bi-scalar AdS invariant distance, which may be expressed in different ways. It is often useful to define the invariant distance $u(x_1,x_2)$ as
\begin{equation} \label{inv_dist}
    u(x_1,x_2)=\frac{(z_1-z_2)^2+(\vec x_1-\vec x_2)^2}{2z_1z_2},
\end{equation}
where we use Poincare coordinates.
From this the chordal distance, $\xi(x_1,x_2)$ (not to be confused with the gauge fixing parameter\footnote{It should hopefully be clear from the context what $\xi$ represents below. The only places where we use $\xi$ as the chordal distance instead of the gauge fixing parameter in the rest of the paper are in equation (\ref{propagator in xi basis}) where we write the gauge field propagator using the chordal distance as the independent variable and in appendix \ref{appendix position space formulas}.}), can be defined as
\begin{equation}\label{cordal}
    \xi=\frac{1}{1+u},
\end{equation}
and the geodesic distance, $\mu(x_1,x_2)$, can be subsequently written as
\begin{equation}
    \mu=\ln\left(\frac{1+\sqrt{1-\xi^2}}{\xi}\right).
\end{equation}
The geodesic distance $\mu$ and $u$ are related by
\begin{equation}
    u=\cosh \mu-1.
\end{equation}
We include in appendix \ref{appendix position space formulas} some useful formulas involving derivatives of these biscalars.

We are now ready to solve for the propagators, which must satisfy equations (\ref{gh eq int}), (\ref{gauge eq int}), and (\ref{BRST eq int}), which we repeat here for convenience (in \eqref{gauge eq sol} and (\ref{BRST eq sol}), $\xi$ is the gauge fixing parameter),
\begin{align}
    \nabla^\mu\nabla_\mu G_{\rm ghost}&=\frac{1}{\sqrt g}\delta^{(d+1)}(x_1-x_2),\label{gh eq sol}\\
    \left(g^{\mu\nu}\nabla^\sigma\nabla_\sigma-\nabla^\mu\nabla^\nu+\frac{1}{\xi}\nabla^\nu\nabla^\mu\right)G_{\mu\rho'}&=-\frac{1}{\sqrt g}\delta^\nu_{\rho'}\delta^{(d+1)}(x_1-x_2),\label{gauge eq sol}\\
    \nabla^\mu G_{\mu\nu'}&=\xi\nabla_{\nu'}G_{\rm ghost}.\label{BRST eq sol}
\end{align}

\subsection{Ghost propagator}

We start with the ghost propagator. It is simplest to work with the geodesic distance $\mu$ as the independent variable. The equation for the ghost propagator then takes the form,
\begin{equation}\label{ghost prop eq in mu variable}
    G_{\rm ghost}''(\mu)+d \coth(\mu)G_{\rm ghost}'(\mu)=\frac{1}{\sqrt g}\delta^{(d+1)}(x_1-x_2).
\end{equation}
Away from coincident points, $x_1\neq x_2$, the solution is given by
\begin{equation}
    G_{\rm ghost}(\mu)=c_{g0} \sech^d(\mu) \, _2F_1\left(\frac{d}{2},\frac{d+1}{2};\frac{d+2}{2};\sech^2(\mu)\right)+c_g.
\end{equation}
The coincidence limit is $\mu \to 0$, and in this limit the solution of \eqref{ghost prop eq in mu variable} should reduce to the flat space propagator,
\begin{equation}
    G_{\rm ghost}^{\rm flat}(\mu)=-\frac{\Gamma\left(\frac{d-1}{2}\right)}{4\pi^{\frac{d+1}{2}}}\frac{1}{\mu^{d-1}}.
\end{equation}
Requiring that to leading order as $\mu\to0$, $G_{\rm ghost}(\mu)$ reduces to $G_{\rm ghost}^{\rm flat}(\mu)$ fixes
\begin{equation}
    c_{0g}=-\frac{\Gamma \left(\frac{d+1}{2}\right)}{2d \pi ^{\frac{d+1}{2}}}.
\end{equation}
Requiring further that the propagator vanishes at large separation, {\it i.e.}\ as $\mu\to\infty$, also fixes $c_g=0$. Thus, the ghost propagator takes the form\footnote{An equivalent form of the solution that makes the behaviour at large separation manifest is 
\begin{equation}
    G_{\rm ghost}(x)=-\frac{\Gamma \left(\frac{d+1}{2}\right)}{2d \pi ^{\frac{d+1}{2}}}(x-1)^{\frac d2}\Big[x^{-d} (x+1)^{\frac d2} \, _2F_1\left(\frac{d}{2},\frac{d+1}{2};\frac{d+2}{2};1-\frac{1}{x^2}\right)\Big],
\end{equation}
where $x=\coth \mu$. The term in square brackets limits to $2^{d/2}$ as $x \to 1$, so the propagator vanishes as $(x-1)^{d/2}$ at large separation. Note also that when $d$ is even this term is proportional to the Jacobi polynomial $P_{(d-2)/2}^{d/2, -d/2}(x)$.}
\begin{equation}\label{covariant gauge ghost propagator in position space}
    G_{\rm ghost}(\mu)=-\frac{\Gamma \left(\frac{d+1}{2}\right)}{2d \pi ^{\frac{d+1}{2}}} \sech^d(\mu) \, _2F_1\left(\frac{d}{2},\frac{d+1}{2};\frac{d+2}{2};\sech^2(\mu)\right).
\end{equation}
We note here that the constant $c_g$ plays no role anyway in the BRST constraint equation (\ref{BRST eq sol}) as only the derivative of the ghost propagator appears.

\subsection{Photon propagator}

To solve for the gauge field propagator we will make the following Ansatz
\begin{equation}
    G_{\mu\rho'}(\mu)=A(\mu)\nabla_\mu\nabla_{\rho'}\mu+B(\mu)\nabla_\mu \mu\nabla_{\rho'}\mu.
\end{equation}
We note here that we have
\begin{equation}
    \nabla^{\mu}\mu \, G_{\mu\rho'}(\mu)=B(\mu)\nabla_{\rho'}\mu.\label{eads transversality}
\end{equation}
We now plug in our Ansatz in equation (\ref{gauge eq sol}) and simplify it for $x_1\neq x_2$ (or equivalently $\mu\neq0$). After some lengthy manipulations we find two second-order coupled differential equations for $A$ and $B$ (where primes on $A$ and $B$ denote derivatives with respect to $\mu$),
\begin{align}
    A''+(d-2) \coth (\mu) A'-\frac{1}{\xi }d A \csch^2(\mu)+\left(\frac{1}{\xi }-1\right) B'-\left(d-2-\frac{d}{\xi }\right) \coth (\mu)B&=0,\label{physeq}\\
    B''+d \coth (\mu) B'-d (1+\xi) B\csch^2(\mu)-d (1-\xi ) \csch^2(\mu) A'+2 d  \coth (\mu) \csch^2(\mu)A&=0.\label{gaugeeq}
\end{align}
Plugging our Ansatz in the BRST constraint equation (\ref{BRST eq sol}) as well, we get
\begin{equation}
    -d \csch^2(\mu)A+B'+d \coth (\mu)B =\xi  G_{\rm ghost}'(\mu).\label{brsteq}
\end{equation}
The derivative of the ghost propagator reduces to a simple function,
\begin{equation}
    G_{\rm ghost}'(\mu)=\frac{\Gamma \left(\frac{d+1}{2}\right)}{2\pi ^{\frac{d+1}{2}}} \csch^d(\mu),
\end{equation}
so we can use equation (\ref{brsteq}) to express $A$ in terms of $B$ as
\begin{equation}
    A=-\xi\frac{ \Gamma \left(\frac{d+1}{2}\right)}{2 d \pi ^{\frac{d+1}{2}}}\csch^{d-2}(\mu)+\frac{1}{d}\sinh ^2(\mu) B'+ \sinh (\mu) \cosh (\mu)B.\label{A in terms of B}
\end{equation}
Plugging this expression of $A$ in terms of $B$ in equation (\ref{gaugeeq}) we find a second-order differential equation for $B$,
\begin{equation}
    B''+(d+2) \coth (\mu) B'+2 d B=(d-2)\left(\frac{d}{d-2}-\xi\right)\frac{\Gamma \left(\frac{d+1}{2}\right)}{2\pi ^{\frac{d+1}{2}}} \cosh (\mu) \text{csch}^{d+1}(\mu).\label{B ode}
\end{equation}
We notice immediately that in the Fried-Yennie gauge the right-hand side is zero and the above reduces to a homogeneous second-order differential equation for $B$, which we can readily solve to get
\begin{equation}
    B=c \sech^d(\mu)\, _2F_1\left(\frac{d}{2},\frac{d+1}{2};\frac{d+3}{2};\tanh ^2(\mu)\right)+c_0 \cosh(\mu)\csch^{d+1}(\mu).
\end{equation}
Using this expression for $B$ we can use equation (\ref{A in terms of B}) to find $A$.  We checked that with these expressions for $A$ and $B$ equation (\ref{physeq}) is automatically satisfied as well. To complete the derivation of the propagator in the Fried-Yennie gauge all that is left to do is to fix the integration constants $c_0$ and $c$. It turns out that unless $c_0=0$, the propagator is more singular than the flat space propagator (given in (\ref{flat space gauge prop in mu variable})) at short distances ({\it i.e.}\ $\mu\to0$), so
\begin{equation}
    B=c \sech^d(\mu)\, _2F_1\left(\frac{d}{2},\frac{d+1}{2};\frac{d+3}{2};\tanh ^2(\mu)\right).
\end{equation}
The propagator should have the fall-off behaviour of normalizable modes as we approach the conformal boundary of EAdS. With Dirichlet boundary conditions this implies that the propagator should vanish as $\mu\to\infty$, {\it i.e.}\ at large separation, and imposing this condition fixes $c=0$. Thus, we finally obtain
\begin{align}\label{form factors in FY gauge}
    A&=-\frac{ \Gamma \left(\frac{d+1}{2}\right)}{2 (d-2) \pi ^{\frac{d+1}{2}}}\frac{1}{\sinh^{d-2}(\mu)},\\
    B&=0.
\end{align}

One may check that the flat space limit is correct: as $\mu\to0$ the gauge field propagator reduces to the flat space propagator given in equation (\ref{flat space gauge prop in mu variable}), including normalisation. Note also that it is sufficient to require that the ghost propagator has the correct flat space limit in order to achieve the same property for the gauge field propagator, precisely because of the BRST constraint (\ref{brsteq}). We also comment that given the flat space expression of the propagator in this basis of bitensors, equation (\ref{flat space gauge prop in mu variable}), it is perhaps unsurprising that a similar result in EAdS holds: since we require that the propagator has the right flat space limit and that it vanishes at large separation, the form factor $B$ should go to zero both for small and large $\mu$, besides satisfying a second-order differential equation. Hence, it may not be that surprising that $B$ turns out to be zero.

Thus, the final answer for the gauge field propagator in the Fried-Yennie gauge is
\begin{equation}
    G_{\mu\rho'}(\mu)=-\frac{ \Gamma \left(\frac{d+1}{2}\right)}{2 (d-2) \pi ^{\frac{d+1}{2}}}\frac{1}{\sinh^{d-2}(\mu)}\nabla_\mu\nabla_{\rho'}\mu,
\end{equation}
and from equation (\ref{eads transversality}) we see that it satisfies the transversality condition,
\begin{equation}
    \nabla^{\mu}\mu \, G_{\mu\rho'}(\mu)=0,
\end{equation}
which is the analogue of the flat space equation (\ref{flat space transversality}).

Using the relations between $\mu$, $u$, and $\xi$, we can also write the propagator in the basis of bitensors formed by these other biscalars,
\begin{align}
    G_{\mu\rho'}(u)&=\frac{\Gamma \left(\frac{d+1}{2}\right)}{2 (d-2)\pi ^{\frac{d+1}{2}}}\left(-\frac{1}{(u (u+2))^{\frac{d-1}{2}}}\nabla_\mu\nabla_{\rho'}u+\frac{u+1}{(u (u+2))^{\frac{d+1}{2}}}\nabla_\mu u\nabla_{\rho'}u\right),\\
    G_{\mu\rho'}(\xi)&=\frac{\Gamma \left(\frac{d+1}{2}\right)}{2 (d-2)\pi ^{\frac{d+1}{2}}}\left(\frac{1}{\xi^2}\left(\frac{\xi^2}{1-\xi^2}\right)^{\frac{d-1}{2}}\nabla_\mu\nabla_{\rho'}\xi+\frac{2 \xi^2-1}{\xi^5}\left(\frac{\xi^2}{1-\xi^2}\right)^{\frac{d+1}{2}}\nabla_\mu \xi\nabla_{\rho'}\xi\right).\label{propagator in xi basis}
\end{align}
Our result written in the $u$-basis agrees with the form given in \cite{Ciccone:2024guw} for Dirichlet boundary conditions.  We include in appendix \ref{appendix position space propagator in covariant gauge for arbitrary xi} the solutions for arbitrary $\xi$ and $d$. Obtaining this most general solution amounts to solving equation (\ref{B ode}) for $B$ for arbitrary $\xi$ and  $d$, deriving an expression for $A$ using equation (\ref{A in terms of B}), and then checking that equation (\ref{physeq}) is also satisfied. As in the discussion so far, we fix the constants of integration by demanding that the propagator reduces to the flat space propagator at short distances and that it vanishes at large distances.
\section{Conclusion}\label{section conclusion}

In this paper, we computed the photon propagator in EAdS for various gauge choices, both in momentum and position space. Our results agree with previous results, whenever such results are available. We made special effort to streamline derivations, and to present results in the simplest possible form. Mathematica notebooks with the results are appended in the arXiv submission to facilitate easy use of our results. Our results straightforwardly extend to non-Abelian gauge theories, as the equations for the propagator only use the quadratic part of the action, which (up to color factors) is the same in Abelian and non-Abelian theories.

We derived and presented our results both in momentum and position space. Each approach has its own advantages, disadvantages, and uses. Momentum space techniques are closer to flat space methodology and, as most explicit QFT computations in flat space are done in momentum space, the results in momentum space would facilitate the transfer of such methodology from flat space to AdS (and indeed the results in \cite{Raju:2011mp} are an example of that). The momentum space expressions are also useful if one is to connect AdS results to CFT in momentum space \cite{Bzowski:2013sza} (via the AdS/CFT correspondence) and also for developing a similar approach for cosmological applications (see, for example, \cite{Bzowski:2023nef,MdAbhishek:2025dhx} and references therein). The momentum space results are simplest in the axial and Coulomb gauges, as these gauge conditions are compatible with the translational isometries along the boundary directions.

Position space expressions, on the other hand, make it easier to make manifest use of AdS isometries, and the results are simpler in covariant $R_\xi$ gauges. The position space propagator is the simplest in the Fried-Yennie gauge, where $\xi=d/(d-2)$. The simplicity of this gauge (in flat, AdS, and dS) is linked with an improved IR behaviour. One may write the propagator in terms of the geodesic distance and in the Fried-Yennie gauge the propagator is ``position-space transverse," meaning that it vanishes when contracted with the geodesic distance vector. The position space propagator would be useful in bulk loop computations, where bulk isometries are used in an essential way, such as in the methodology developed in \cite{Banados:2022nhj,BIStoappear}.

In our analysis we paid special attention to the subsidiary condition that BRST invariance imposes: the longitudinal part of the gauge field propagator is linked with the propagator of the ghost field. This relation is instrumental in proving perturbative unitarity in flat space and its importance in de Sitter was also recently emphasised in \cite{Glavan:2022nrd}. In our case, the subsidiary condition has proved very useful when solving for the gauge field propagator, as the equations one gets from the subsidiary condition are simpler than the original equations for the propagator. 
 
We derived the propagator for the standard Dirichlet boundary conditions. This boundary condition implies in particular that the propagator vanishes at large separation (the propagator should fall off at the rate of normalizable modes). In AdS, other boundary conditions for gauge fields have been discussed in the literature, see for example \cite{Witten:2003ya,Marolf:2006nd}. With Neumann boundary conditions one expects dynamical gauge fields on the boundary and the propagator should not vanish at large separation. For Abelian theories (and in $d=3$), Neumann boundary conditions are related to Dirichlet boundary conditions by electromagnetic duality \cite{Witten:2003ya}. For non-Abelian gauge theories, Neumann boundary conditions have been discussed in the context of confinement in AdS \cite{Ciccone:2024guw,Aharony:2012jf}. It would be interesting to extend the analysis presented here to general boundary conditions. 

Another interesting extension is to compute the propagator for other spinning fields in (A)dS. The next (and perhaps most interesting) case is the graviton propagator. In this case, the ghost fields are vector fields and there is a 2-parameter generalisation of the $R_\xi$-type covariant gauges. For these gauges, the ghost action is akin to that of a massive spin-1 field. Similarly to the gauge field case discussed here, BRST invariance couples the longitudinal part of the graviton propagator to the ghost propagator and the methodology we developed here can be used to solve for the propagators. Higher spin fields and $p$-forms are other possible extensions. We hope to report on such extensions in future work.

\section*{Acknowledgements}

We would like to acknowledge Chandramouli Chowdhury and Ioannis Matthaiakakis for useful discussions throughout the development of this project. We also thank Riccardo Ciccone for discussions and correspondence. RM is supported by an STFC studentship. KS is supported in part by the STFC consolidated grant ST/X000583/1 ``New Frontiers in Particle Physics, Cosmology and Gravity."
\appendix
\section{Flat space propagators}\label{appendix flat space propagators}

In this appendix we confirm the BRST constraint (\ref{brst constraint}) between the gauge field and ghost propagators in Euclidean flat space $g_{\mu\nu}=\delta_{\mu\nu}=\rm diag (1,\ldots,1)$ for various well-known gauges.

\subsection{Axial gauge}

Choosing $F(A)=n^\mu A_\mu$, in flat space the action (\ref{action}) becomes
\begin{equation}
    S=\int \mathrm{d}^{d+1}x \left[\frac14 F^{\mu\nu}F_{\mu\nu}+\frac{1}{2\xi}\left(n^\mu A_\mu\right)^2+b n^\mu\partial_\mu c\right].
\end{equation}
Going to momentum space, the photon and ghost propagators must satisfy
\begin{align}
    \left(-p^2\delta_{\mu\nu}+p_\mu p_\nu-\frac{1}{\xi}n_\mu n_\nu\right)\tilde G^{\mu\rho}(p)&=-\delta_\nu^\rho,\\
    \left(-in^\mu p_\mu\right)\tilde G_{\rm ghost}(p)&=1.
\end{align}
From these, we find that
\begin{align}
    \tilde G_{\mu\rho}(p)&=\frac{1}{p^2}\left(\delta_{\mu\rho}-\frac{p_\mu n_\rho+n_\mu p_\rho}{(n\cdot p)}+\frac{\xi p^2+n^2}{(n\cdot p)^2}p_\mu p_\rho\right),\\
    \tilde G_{\rm ghost}(p)&=\frac{i}{n\cdot p}.
\end{align}
Note that
\begin{equation}
    n^\mu\tilde G_{\mu\rho}(p)=\frac{\xi p_\rho}{n\cdot p}=\xi(ip_\rho)\frac{i}{n\cdot(-p)}=\xi (ip_\rho)\tilde G_{\rm ghost}(-p),
\end{equation}
which is precisely the BRST constraint (\ref{brst constraint}).

\subsection{Coulomb gauge}

Choosing $F(A)=\partial^i A_i$ (the sum is over only $d$ of the $(d+1)$ components; since we are in Euclidean signature it does not matter which component is omitted), in flat space the action (\ref{action}) becomes
\begin{equation}
    S=\int \mathrm{d}^{d+1}x \left[\frac14 F^{\mu\nu}F_{\mu\nu}+\frac{1}{2\xi}\left(\partial^i A_i\right)^2+b \partial^i\partial_i c\right].
\end{equation}
Going to momentum space, the photon and ghost propagators must satisfy
\begin{align}
    \left(-p^2\delta_{\mu\nu}+p_\mu p_\nu-\frac{1}{\xi}\delta_\mu^i\delta_\nu^j p_i p_j\right)\tilde G^{\mu\rho}(p)&=-\delta_\nu^\rho,\\
    \left(-p^i p_i\right)\tilde G_{\rm ghost}(p)&=1.
\end{align}
We define $n^\mu=(1,\vec 0)$ and $p_s^\mu=p^\mu-(n\cdot p) n^\mu$. From these, we find that
\begin{align}
    \tilde G_{\mu\rho}(p)=\frac{1}{p_s^2}\Bigg(&n_\mu n_\rho+\frac{p_s^2}{p^2}\left((\delta_{\mu\rho}-n_\mu n_\rho)-\frac{(p_s)_\mu(p_s)_\rho}{p_s^2}\right)\\
    &+\xi\frac{(n\cdot p)^2}{p_s^2}n_\mu n_\rho+\xi\frac{(n\cdot p)\left((p_s)_\mu n_\rho+n_\mu (p_s)_\rho\right)}{p_s^2}+\xi\frac{(p_s)_\mu(p_s)_\rho}{p_s^2}\Bigg),\nonumber\\
    \tilde G_{\rm ghost}(p)=-\frac{1}{p_s^2}&.
\end{align}
Note that
\begin{equation}
    -ip_s^\mu\tilde G_{\mu\rho}(p)=-ip^i\tilde G_{i\rho}(p)=\xi\frac{-ip_\rho}{p_s^2}=\xi(ip_\rho)\left(-\frac{1}{p_s^2}\right)=\xi (ip_\rho)\tilde G_{\rm ghost}(-p),
\end{equation}
which is precisely the BRST constraint (\ref{brst constraint}).

\subsection{Covariant gauge}

Choosing $F(A)=\partial^\mu A_\mu$, in flat space the action (\ref{action}) becomes
\begin{equation}
    S=\int \mathrm{d}^{d+1}x \left[\frac14 F^{\mu\nu}F_{\mu\nu}+\frac{1}{2\xi}\left(\partial^\mu A_\mu\right)^2+b \partial^\mu\partial_\mu c\right].
\end{equation}
Going to momentum space, the photon and ghost propagators must satisfy
\begin{align}
    \left(-p^2\delta_{\mu\nu}+p_\mu p_\nu-\frac{1}{\xi}p_\mu p_\nu\right)\tilde G^{\mu\rho}(p)&=-\delta_\nu^\rho,\\
    \left(-p^\mu p_\mu\right)\tilde G_{\rm ghost}(p)&=1.
\end{align}
From these, we find that
\begin{align}
    \tilde G_{\mu\rho}(p)&=\frac{1}{p^2}\left(\delta_{\mu\rho}-\left(1-\xi\right)\frac{p_\mu p_\rho}{p^2}\right),\label{covariant flat space prop}\\
    \tilde G_{\rm ghost}(p)&=-\frac{1}{p^2}.
\end{align}
Note that
\begin{equation}
    -ip^\mu\tilde G_{\mu\rho}(p)=\xi\frac{-ip_\rho}{p^2}=\xi(ip_\rho)\left(-\frac{1}{p^2}\right)=\xi (ip_\rho)\tilde G_{\rm ghost}(-p),
\end{equation}
which is precisely the BRST constraint (\ref{brst constraint}).
\section{Integral representation of the propagators}\label{appendix integral representation}

In this appendix we give an integral representation for the propagators that avoids the use of Heaviside step functions in their definitions. We will make use of the formulas \cite{Gradshteyn:1702455}
\begin{align}
    \int_0^\infty xJ_\nu(ax)J_\nu(bx)\mathrm{d}x&=\frac{1}{a}\delta(a-b), \quad a,b,\nu\in \mathbb{R},\label{formula 1}\\
    \int_0^\infty \frac x{x^2+c^2}J_\nu(ax)J_\nu(bx)\mathrm{d}x&=I_{\nu}(c b)K_{\nu}(c a)\Theta(a-b)+K_{\nu}(c b)I_{\nu}(c a)\Theta(b-a),\label{formula 2}\\
    &\quad\quad\quad\quad\quad\quad\quad\quad\quad\quad\quad\quad\quad\quad a,b>0, \, \rm Re (c)>0, \, \rm Re (\nu)>-1,\nonumber\\
    \int_0^\infty \frac 1x J_\nu(ax)J_\nu(bx)\mathrm{d}x&=\frac{1}{2\nu}\left(\frac{b}{a}\right)^{\nu}\Theta(a-b)+\frac{1}{2\nu}\left(\frac{a}{b}\right)^{\nu}\Theta(b-a), \quad a,b>0, \, \rm Re (\nu)>0.\label{formula 3}
\end{align}
Note that (\ref{formula 3}) is just (\ref{formula 2}) with $c=0$.

\subsection{Gauge independent part of the propagator}

Using (\ref{formula 2}), we can immediately rewrite the transverse, gauge independent, part of the propagator as
\begin{equation}
    \begin{aligned}
        A&=\left((zz')^{\frac {d-2}2}I_{\frac{d-2}{2}}(p z')K_{\frac{d-2}{2}}(p z)\right)\Theta(z-z')+\left((zz')^{\frac {d-2}2}K_{\frac{d-2}{2}}(p z')I_{\frac{d-2}{2}}(p z)\right)\Theta(z'-z)\\
        &=\int_0^\infty \frac {k \, dk}{k^2+p^2}(z)^\frac{d-2}{2}J_{\frac{d-2}{2}}(kz)J_{\frac{d-2}{2}}(kz')(z')^\frac{d-2}{2}
    \end{aligned}.\label{int rep of A}
\end{equation}

\subsection{Coulomb gauge propagator}

Using (\ref{int rep of A}) and (\ref{formula 1}), we can rewrite the propagator in Coulomb gauge, (\ref{coulomb gauge propagator}), as
\begin{equation}
    \begin{cases}
        \tilde{G}_{ij}(z,z',\vec p)&=\left(\delta_{ij}-\frac{p_ip_j}{p^2}\right)\int_0^\infty \frac {k \, dk}{k^2+p^2}(z)^\frac{d-2}{2}J_{\frac{d-2}{2}}(kz)J_{\frac{d-2}{2}}(kz')(z')^\frac{d-2}{2}\\
        \tilde{G}_{i0}(z,z',\vec p)&=\tilde{G}_{0i}(z,z',\vec p)=0\\
        \tilde{G}_{00}(z,z',\vec p)&=\frac{z^{d-3}}{p^2}\delta(z-z')=\int_0^\infty \frac{k \, dk}{p^2}(z)^\frac{d-2}{2}J_{\frac{d-2}{2}}(kz)J_{\frac{d-2}{2}}(kz')(z')^\frac{d-2}{2}
    \end{cases}.
\end{equation}
(For the $00$-component we could have used (\ref{formula 1}) with any order of the Bessel J.)

\subsection{Axial gauge propagator}

For convenience, we rewrite here the propagator in axial gauge from (\ref{axial gauge propagator}),
\begin{equation}
    \begin{cases}
        \tilde{G}_{ij}(z,z',\vec p)&=\left(\delta_{ij}-\frac{p_ip_j}{p^2}\right)\begin{cases}
        (zz')^{\frac {d-2}2}K_{\frac{d-2}{2}}(p z')I_{\frac{d-2}{2}}(p z), \quad z<z'\\
        (zz')^{\frac {d-2}2}I_{\frac{d-2}{2}}(p z')K_{\frac{d-2}{2}}(p z), \quad z>z'
    \end{cases}+\frac{p_ip_j}{p^2}\begin{cases}
        \frac{1}{d-2}z^{d-2}, \quad z<z'\\
        \frac{1}{d-2}z'^{d-2}, \quad z>z'
    \end{cases}\\
        \tilde{G}_{i0}(z,z',\vec p)&=\tilde{G}_{0i}(z,z',\vec p)=\tilde{G}_{00}(z,z',\vec p)=0
    \end{cases}.
\end{equation}
The longitudinal component can be rewritten as
\begin{equation}
    \begin{aligned}
        \frac{p_ip_j}{p^2}\begin{cases}
            \frac{1}{d-2}z^{d-2}, \quad z<z'\\
            \frac{1}{d-2}z'^{d-2}, \quad z>z'
        \end{cases}&=\frac{p_ip_j}{p^2}(zz')^{\frac{d-2}{2}}\left(\frac1{d-2}\left(\frac{z'}{z}\right)^{\frac{d-2}{2}}\Theta(z-z')+\frac1{d-2}\left(\frac{z}{z'}\right)^{\frac{d-2}{2}}\Theta(z'-z)\right)\\
        &=\frac{p_ip_j}{p^2}\int_0^\infty \frac{dk}{k} (z)^{\frac{d-2}{2}}J_{\frac{d-2}{2}}(kz)J_{\frac{d-2}{2}}(kz')(z')^{\frac{d-2}{2}},
    \end{aligned}
\end{equation}
where we used (\ref{formula 3}) to get the second line. With this, the axial gauge propagator can be rewritten as
\begin{equation}
    \begin{cases}
        \tilde{G}_{ij}(z,z',\vec p)&=\left(\delta_{ij}-\frac{p_ip_j}{p^2}\right)\int_0^\infty \frac {k \, dk}{k^2+p^2}(z)^\frac{d-2}{2}J_{\frac{d-2}{2}}(kz)J_{\frac{d-2}{2}}(kz')(z')^\frac{d-2}{2}\\
        &\quad +\frac{p_ip_j}{p^2}\int_0^\infty \frac{dk}{k} (z)^{\frac{d-2}{2}}J_{\frac{d-2}{2}}(kz)J_{\frac{d-2}{2}}(kz')(z')^{\frac{d-2}{2}}\\
        &=\int_0^\infty \frac {k \, dk}{k^2+p^2}(z)^\frac{d-2}{2}J_{\frac{d-2}{2}}(kz)J_{\frac{d-2}{2}}(kz')(z')^\frac{d-2}{2}\left(\delta_{ij}+\frac{p_ip_j}{k^2}\right)\\
        \tilde{G}_{i0}(z,z',\vec p)&=\tilde{G}_{0i}(z,z',\vec p)=\tilde{G}_{00}(z,z',\vec p)=0
    \end{cases}.
\end{equation}
Up to some factors of $i$ due to the different signature of spacetime, this agrees with the expression given in \cite{Raju:2011mp}.

\subsection{Axial and Coulomb gauge propagators in position space}

The position space propagators are given by
\begin{equation}
    G_{\mu\nu}(x,y)=\int\frac{\mathrm{d}^dp}{(2\pi)^d}\tilde{G}_{\mu\nu}(z,z',\vec p)e^{-i\vec p\cdot(\vec x-\vec y)}.
\end{equation}

\subsubsection*{Axial gauge}

The axial gauge propagator in position space is given by
\begin{equation}
G_{ij}^{\rm axial}(x,y)= \delta_{ij} A(x, y) + \partial_{x_i}\partial_{y_j} B(x, y)\, ,    
\end{equation}
where the form factors $A$ and $B$ can be computed using
\begin{equation}
    G_{ij}^{\rm axial}(x,y)=(zz')^{\frac{d-2}{2}}\int_0^\infty dk\,k\,J_{\frac{d-2}{2}}(kz)J_{\frac{d-2}{2}}(kz')\left(\delta_{ij}+\frac{\partial_{x_i}\partial_{y_j}}{k^2}\right)\int\frac{\mathrm{d}^dp}{(2\pi)^d}\frac1{p^2+k^2}e^{-i\vec p\cdot(\vec x-\vec y)}.
\end{equation}
The integral over $\vec p$ gives just the flat space position space propagator of a scalar with mass $k$, and it can be computed by introducing a Schwinger parameter, completing the square and integrating over $\vec p$, and then performing the integral over the Schwinger parameter. The result is
\begin{equation}
    \int\frac{\mathrm{d}^dp}{(2\pi)^d}\frac1{p^2+k^2}e^{-i\vec p\cdot(\vec x-\vec y)}=\frac1{(2\pi)^{\frac d2}}\frac{1}{|\vec x-\vec y|^{\frac{d-2}{2}}}k^{\frac{d-2}{2}}K_{\frac{d-2}{2}}(k|\vec x-\vec y|).
\end{equation}
Thus, the propagator is given by
\begin{equation}
    \begin{aligned}
        G_{ij}^{\rm axial}(x,y)=\frac{(zz')^{\frac{d-2}{2}}}{(2\pi)^{\frac d2}}\Bigg[\delta_{ij}&\frac{1}{|\vec x-\vec y|^{\frac{d-2}{2}}}\int_0^\infty dk\,k^{\frac d2}\,J_{\frac{d-2}{2}}(kz)J_{\frac{d-2}{2}}(kz')K_{\frac{d-2}{2}}(k|\vec x-\vec y|)+\\
        +\partial_{x_i}\partial_{y_j}&\frac{1}{|\vec x-\vec y|^{\frac{d-2}{2}}}\int_0^\infty dk\,k^{\frac d2-2}\,J_{\frac{d-2}{2}}(kz)J_{\frac{d-2}{2}}(kz')K_{\frac{d-2}{2}}(k|\vec x-\vec y|)\Bigg].
    \end{aligned}
\end{equation}
The two integrals can be written in terms of Appell $F_4$ functions \cite{Gradshteyn:1702455} as
\begin{align}
    \int_0^\infty dk\,k^{\frac d2}\,J_{\frac{d-2}{2}}(kz)J_{\frac{d-2}{2}}(kz')K_{\frac{d-2}{2}}(k|\vec x-\vec y|)&=\frac{2^{\frac{d-2}{2}}\Gamma\left(d-1\right)}{\Gamma\left(\frac d2\right)}\frac{(zz')^{\frac{d-2}{2}}}{|\vec x-\vec y|^{\frac{3d-2}{2}}} \label{JJK} \\
    &\quad \times F_4\left(\frac d2,d-1;\frac d2,\frac d2;\frac{-z^2}{|\vec x-\vec y|^2},\frac{-z'^2}{|\vec x-\vec y|^2}\right)\nonumber \\
    &=\frac{2^{\frac{3d-6}{2}}\Gamma\left(\frac{d-1}{2}\right)(zz'|\vec x-\vec y|)^{\frac{d-2}{2}}}{\sqrt \pi \left[(z^2+z'^2+|\vec x-\vec y|^2)^2-4z^2z'^2\right]^{\frac{d-1}{2}}}\, , \nonumber\\
    \int_0^\infty dk\,k^{\frac d2-2}\,J_{\frac{d-2}{2}}(kz)J_{\frac{d-2}{2}} (kz')K_{\frac{d-2}{2}}(k|\vec x-\vec y|)&=\frac{2^{\frac{d-4}{2}}\Gamma\left(d-2\right)}{(d-2)\Gamma\left(\frac d2\right)}\frac{(zz')^{\frac{d-2}{2}}}{|\vec x-\vec y|^{\frac{3d-6}{2}}} \label{JJK2} \\
    &\quad \times F_4\left(\frac{d-2}{2},d-2;\frac d2,\frac d2;\frac{-z^2}{|\vec x-\vec y|^2},\frac{-z'^2}{|\vec x-\vec y|^2}\right).\nonumber
\end{align}
The integral in \eqref{JJK2} may be computed in terms of elementary functions when $d$ is odd (the Bessel functions reduce to elementary functions in such cases), and for any $d$ one may be able to connect this integral to \eqref{JJK} using the methodology in \cite{Bzowski:2015yxv}. 

The final answer for $A(x, y)$ is then given by 
\begin{equation}
A(x, y) = \frac{\Gamma\left(\frac{d-1}{2}\right) \xi^{d-1}}{4\pi^{\frac{d+1}{2}}(zz')\left(1 -{\xi^2}\right)^{\frac{d-1}{2}}}=\frac{\Gamma\left(\frac{d-1}{2}\right)}{4\pi^{\frac{d+1}{2}}(zz')\left(u(u+2)\right)^{\frac{d-1}{2}}},
\end{equation}
where $u$ and $\xi$ are the AdS invariant and chordal distances defined in \eqref{inv_dist} and \eqref{cordal}. This factor exactly reproduces the gauge invariant part of the propagator computed in the covariant gauge in \cite{DHoker:1998bqu}. To check this one may use their (2.8), (2.6), and (2.18).

\subsubsection*{Coulomb gauge}

The Coulomb gauge propagator in position space can be obtained in the same way. We get
\begin{equation} \label{00_Coulomb}
    G_{00}^{\rm Coulomb}(x,y)=z^{d-3}\delta(z-z')\int\frac{\mathrm{d}^dp}{(2\pi)^d}\frac1{p^2}e^{-i\vec p\cdot(\vec x-\vec y)}=\frac{\Gamma\left(\frac{d-2}{2}\right)}{4\pi^{\frac d2}}\frac{1}{|\vec x-\vec y|^{d-2}}z^{d-3}\delta(z-z')
\end{equation}
and
\begin{equation}
    \begin{aligned}
        G_{ij}^{\rm Coulomb}(x,y)&=G_{ij}^{\rm axial}(x,y)\\
        &-\frac{\Gamma\left(\frac{d-2}{2}\right)}{4(d-2)\pi^{\frac d2}}\partial_{x_i}\partial_{y_j}\frac{1}{|\vec x-\vec y|^{d-2}}\left(z'^{d-2}\Theta(z-z')+z^{d-2}\Theta(z'-z)\right).
    \end{aligned}
\end{equation}
\section{Some useful formulas in position space}\label{appendix position space formulas}

We include here some useful formulas in position space involving the biscalars $u$, $\xi$, and $\mu$ defined at the beginning of section \ref{section position space 2} and their derivatives. We are working in EAdS$_{d+1}$, so the spacetime dimension is $D=d+1$. We recall that $u(x_1,x_2)$ is defined as
\begin{equation}
    u(x_1,x_2)=\frac{(z_1-z_2)^2+(\vec x_1-\vec x_2)^2}{2z_1z_2},
\end{equation}
and we have that $u\ge0$. From equations (2.6-2.7) of \cite{DHoker:1998bqu} we have
\begin{align}
    \nabla_\mu u&=\frac1{z_1}\left[\frac1{z_2}(x_1-x_2)_\mu-u\delta_{\mu0}\right],\\
    \nabla_{\nu'}u&=\frac1{z_2}\left[\frac1{z_1}(x_2-x_1)_{\nu'}-u\delta_{\nu'0}\right],\\
    \nabla_\mu \nabla_{\nu'}u&=-\frac1{z_1z_2}\left[\delta_{\mu\nu'}+\frac1{z_2}(x_1-x_2)_\mu\delta_{\nu'0}+\frac1{z_1}(x_2-x_1)_{\nu'}\delta_{\mu0}-u\delta_{\mu0}\delta_{\nu'0}\right],
\end{align}
and from equations (2.9-2.15) of \cite{DHoker:1999bve} we have
\begin{align}
    \nabla^\mu\nabla_\mu u&=(d+1)(u+1),\\
    \nabla^\mu u \, \nabla_\mu u&=u(u+2),\\
    \nabla_\mu \nabla_\nu u&=(u+1)g_{\mu\nu},\\
    (\nabla^\mu u)(\nabla_\mu \nabla_\nu \nabla_{\nu'} u)&=\nabla_\nu u \, \nabla_{\nu'} u,\\
    (\nabla^\mu u)(\nabla_\mu \nabla_{\nu'} u)&=(u+1)\nabla_{\nu'}u,\\
    (\nabla^\mu \nabla_{\mu'}u)(\nabla_\mu \nabla_{\nu'}u)&=g_{\mu' \nu'}+\nabla_{\mu'} u \, \nabla_{\nu'} u,\\
    \nabla_\mu \nabla_\nu \nabla_{\nu'}u&=g_{\mu\nu}\nabla_{\nu'}u.
\end{align}
For any scalar function $f$ of $u$ we have
\begin{equation}
    \nabla^\mu\nabla_\mu f(u)=u(u+2)f''(u)+(d+1)(u+1)f'(u).
\end{equation}

From $u$ we can define the chordal distance $\xi(x_1,x_2)$ as
\begin{equation}
    \xi=\frac{1}{1+u},
\end{equation}
which satisfies $0<\xi\le1$. If $f$ is viewed as a scalar function of $\xi$ now, we have
\begin{equation}
    \nabla^\mu\nabla_\mu f(\xi)=(1-\xi^2)\xi^2 f''(\xi)+[2(1-\xi^2)-(d+1)]\xi f'(\xi).
\end{equation}
Using the relation between $u$ and $\xi$, one can derive that
\begin{align}
    \nabla_\mu u&=-\frac{1}{\xi^2}\nabla_\mu \xi,\\
    \nabla_\mu\nabla_{\nu'}u&=-\frac{1}{\xi^2}\nabla_\mu\nabla_{\nu'}\xi+\frac{2}{\xi^3}\nabla_\mu\xi\nabla_{\nu'}\xi.
\end{align}

The geodesic distance $\mu(x_1,x_2)$ can be written in terms of $\xi$ as
\begin{equation}
    \mu=\ln\left(\frac{1+\sqrt{1-\xi^2}}{\xi}\right),
\end{equation}
and it is related to $u$ by
\begin{equation}
    u=\cosh \mu-1.
\end{equation}
We also have that $\mu\ge0$. Using the relation between $u$ and $\mu$, one can derive that
\begin{align}
    \nabla_\mu u&=\sinh \mu \, \nabla_\mu \mu,\\
    \nabla_\mu\nabla_{\nu'}u&=\sinh \mu \, \nabla_\mu\nabla_{\nu'}\mu+\cosh \mu \, \nabla_\mu\mu\nabla_{\nu'}\mu.
\end{align}

Finally, using the relations above for the derivatives of $u$, one can also get
\begin{align}
    \nabla^\mu\mu\nabla_\mu\mu&=1,\\
    \nabla^\mu\nabla_\mu\mu&=d \coth \mu,\\
    \nabla_\mu\nabla_\nu\mu&=\coth \mu \, (g_{\mu\nu}-\nabla_\mu\mu\nabla_\nu\mu),\\
    \nabla^\mu\mu\nabla_\mu\nabla_\nu\mu&=0,\\
    \nabla_\mu\nabla_\nu\nabla_{\nu'}\mu&=-\coth \mu \, (\nabla_\nu\mu\nabla_\mu\nabla_{\nu'}\mu+\nabla_\mu\mu\nabla_\nu\nabla_{\nu'}\mu)-\csch^2\mu \, (g_{\mu\nu}-\nabla_\mu\mu\nabla_\nu\mu)\nabla_{\nu'}\mu,\\
    \nabla^\mu\mu\nabla_\mu\nabla_{\nu'}\mu&=0,\\
    \nabla^\mu\mu\nabla_\mu\nabla_\nu\nabla_{\nu'}\mu&=-\coth\mu \, \nabla_\nu\nabla_{\nu'}\mu,\\
    \nabla^\mu\nabla_\mu\nabla_\nu\nabla_{\nu'}\mu&=2d\coth\mu\csch^2\mu \, \nabla_\nu\mu\nabla_{\nu'}\mu-d\coth^2\mu \, \nabla_\nu\nabla_{\nu'}\mu,\\
    \nabla^\mu\nabla_\mu\nabla_\nu\mu&=-d\coth^2\mu \, \nabla_\nu\mu,\\
    \nabla^\mu\nabla_\mu\nabla_{\nu'}\mu&=-d\csch^2\mu \, \nabla_{\nu'}\mu.
\end{align}
\section{Position space propagator in covariant gauge for arbitrary $\xi$}\label{appendix position space propagator in covariant gauge for arbitrary xi}

\vspace{0.5cm}

\begin{equation}
    G_{\mu\rho'}(\mu)=A(\mu)\nabla_\mu\nabla_{\rho'}\mu+B(\mu)\nabla_\mu \mu\nabla_{\rho'}\mu
\end{equation}

\vspace{0.5cm}

\subsection{$d=3$}

\begin{align}
    G_{\rm ghost}(\mu)&=\frac{1}{4 \pi ^2}\left(\tanh ^{-1}(\text{sech}(\mu ))-\coth (\mu ) \text{csch}(\mu )\right)\\
    A(\mu)&=-\frac{\text{csch}(\mu)}{2 \pi ^2}+\frac{\xi-3}{48\pi^2}\text{csch}^3(\mu)\times\\
    &\times \left(-2 \cosh (2 \mu)+8 \log (2 \text{csch}(\mu))+(\cosh (3 \mu)-9 \cosh (\mu)) \tanh ^{-1}(\text{sech}(\mu))+2\right)\nonumber\\
    B(\mu)&=\frac{\xi-3}{8 \pi ^2} \text{csch}^4(\mu ) \left(\tanh ^{-1}(\text{sech}(\mu )) (3+\cosh (2 \mu ))+4
   \cosh (\mu ) \log \left(\frac{\sinh (\mu )}{2}\right)\right)
\end{align}

\vspace{0.5cm}

\subsection{$d=4$}

\begin{align}
    G_{\rm ghost}(\mu)&=-\frac{1}{8 \pi ^2}\left(\coth ^3(\mu)-3 \coth (\mu)+2\right)\\
    A(\mu)&=-\frac{3\, \text{csch}^2(\mu )}{16 \pi ^2}+\frac{\xi-2}{64\pi^2}\text{csch}^4(\mu )\times\\
    &\times (\sinh (2 \mu )-\cosh (2 \mu )) ((6 \mu -3) \sinh (2 \mu )+(6 \mu -4) \cosh (2 \mu )+4)\nonumber\\
    B(\mu)&=\frac{\xi-2}{8 \pi ^2} \text{csch}^2(\mu ) \left(-1+\coth (\mu )+3 (-1+\mu  \coth (\mu ))
   \text{csch}^2(\mu )\right)
\end{align}

\vspace{0.5cm}

\subsection{$d=5$}

\begin{align}
    G_{\rm ghost}(\mu)&=\frac{1}{8 \pi^3}\left(\coth (\mu ) \text{csch}(\mu ) \left(3-2 \text{csch}^2(\mu )\right)-3 \tanh ^{-1}(\text{sech}(\mu ))\right)\\
    A(\mu)&=-\frac{\text{csch}^3(\mu )}{3 \pi ^3}-\frac{3\xi-5}{1920\pi^3}\text{csch}^5(\mu )\times\nonumber\\
    &\times \Big(48 \cosh (2 \mu )-6 \cosh(4 \mu )-128 \log (2 \text{csch}(\mu ))-42\\
    &+(150 \cosh (\mu )-25
   \cosh (3 \mu )+3 \cosh (5 \mu )) \tanh ^{-1}(\text{sech}(\mu ))\Big)\nonumber\\
    B(\mu)&=\frac{3 \xi -5}{192 \pi ^3} \text{csch}^6(\mu )\times\nonumber\\
    &\times\Big(\tanh ^{-1}(\text{sech}(\mu )) (45+20 \cosh (2 \mu
   )-\cosh (4 \mu ))\\
   &+4 \cosh (\mu ) (-1+\cosh (2 \mu )-16 \log (2)+16 \log (\sinh (\mu )))\Big)\nonumber
\end{align}

\vspace{0.5cm}

\subsection{$d=6$}

\begin{align}
    G_{\rm ghost}(\mu)&=\frac{1}{32 \pi ^3}\left(\text{csch}^5(\mu ) (9 \sinh (2 \mu )+11 \cosh (2 \mu )-5) (\sinh (3 \mu )-\cosh (3 \mu ))\right)\\
    A(\mu)&=-\frac{15 \,\text{csch}^4(\mu )}{64 \pi ^3}-\frac{2\xi-3}{192\pi^3}\times\nonumber\\
    &\times\Big(15\mu\,  \text{csch}^6(\mu )-6\, \text{csch}^2(\mu )+8\\
    &+\text{csch}^4(\mu ) (9-(-9 \cosh (2 \mu )+\cosh (4 \mu )+23) \coth
   (\mu ))\Big)\nonumber\\
    B(\mu)&=\frac{2 \xi-3}{32 \pi ^3} \text{csch}^2(\mu ) \times\\
    &\times\left(2-2 \coth (\mu )+(-5+6 \coth (\mu ))
   \,\text{csch}^2(\mu )+15 (-1+\mu  \coth (\mu )) \,\text{csch}^4(\mu )\right)\nonumber
\end{align}

\subsection{General $d$}

Obtaining a solution for the propagator for arbitrary $d$ and $\xi$ boils down to solving equation (\ref{B ode}),
\begin{equation}
    B''+(d+2) \coth (\mu) B'+2 d B=(d-2)\left(\frac{d}{d-2}-\xi\right)\frac{\Gamma \left(\frac{d+1}{2}\right)}{2\pi ^{\frac{d+1}{2}}} \cosh (\mu) \text{csch}^{d+1}(\mu)\equiv S(\mu)\, ,\label{B ode appendix}
\end{equation}
where we denote the right-hand side, {\it i.e.}\ the source term, simply by $S$. Mathematica cannot solve this differential equation directly, however, since the homogeneous solution corresponding to the Fried-Yennie gauge has been found, one may solve for the particular integral using the method of variation of parameters. The two independent solutions of the homogeneous equation are given by
\begin{align}
    B_1&=\cosh(\mu)\csch^{d+1}(\mu),\\
    B_2&=\sech^d(\mu)\, _2F_1\left(\frac{d}{2},\frac{d+1}{2};\frac{d+3}{2};\tanh ^2(\mu)\right).
\end{align}
The Wronskian is
\begin{equation}
    W=\csch(\mu)^{d+2}.
\end{equation}
A particular integral for the differential equation can be constructed from those as follows,
\begin{equation}
    B_{\rm PI}=B_1b_1+B_2b_2,
\end{equation}
where $b_1$ and $b_2$ are given by
\begin{align}
    b_1&=-\int\frac{1}{W}B_2S\,\mathrm{d}\mu=(d-2) \left(\xi-\frac{d}{d-2}\right)\frac{\Gamma\left(\frac{d+1}{2}\right)}{2\pi ^{\frac{d+1}{2}}}\times\\ &\quad\quad\quad\quad\quad\quad\quad\quad\,\,\times\int\sinh (\mu) \text{sech}^{d-1}(\mu) \,_2F_1\left(\frac{d}{2},\frac{d+1}{2};\frac{d+3}{2};\tanh ^2(\mu)\right)\,\mathrm{d}\mu,\nonumber\\
    b_2&=\int\frac{1}{W}B_1S\,\mathrm{d}\mu=(d-2) \left(\frac{d}{d-2}-\xi\right)\frac{\Gamma\left(\frac{d+1}{2}\right)}{2\pi ^{\frac{d+1}{2}}}\int\cosh ^2(\mu) \text{csch}^d(\mu)\,\mathrm{d}\mu.
\end{align}
With the substitution $t=\tanh^2(\mu)$, $0\le t<1$, the two integrals above become
\begin{align}
    I_1=\frac12\int&(1-t)^{\frac{d-4}{2}}\, _2F_1\left(\frac{d}{2},\frac{d+1}{2};\frac{d+3}{2};t\right)\mathrm{d}t,\\
    I_2=\frac12\int&(1-t)^{\frac{d-4}{2}}\, t^{-\frac{d+1}{2}}\mathrm{d}t.
\end{align}

\subsubsection*{Generic $d$}

The first integral gives
\begin{equation}
    I_1=-\frac{3t}{2(d-2)} \, _3F_2\left(1,1,\frac{5}{2};2,\frac{d+3}{2};t\right)-\frac{d+1}{2(d-2)} \log (1-t).
\end{equation}
The second integral, for generic values of $d$, gives
\begin{equation}
    I_2=-\frac{1}{d-1}t^{-\frac{d-1}{2}} \, _2F_1\left(\frac{1-d}{2},\frac{4-d}{2};\frac{3-d}{2};t\right)+\frac{1}{4 \sqrt{\pi }}\Gamma \left(\frac{1-d}{2}\right) \Gamma \left(\frac{d-2}{2}\right).
\end{equation}
This, however, breaks down when $d$ is an odd integer. We will deal with this case separately. (The constants of integration have been chosen for later convenience.)

So, as long as $d$ is not an odd integer, the full solution to (\ref{B ode appendix}) is given by
\begin{equation}
    \begin{aligned}
        B&=c \sech^d(\mu)\, _2F_1\left(\frac{d}{2},\frac{d+1}{2};\frac{d+3}{2};\tanh ^2(\mu)\right)+c_0 \cosh(\mu)\csch^{d+1}(\mu)\\
        &+\left(d-(d-2)\xi\right)\frac{\Gamma\left(\frac{d+1}{2}\right)}{2\pi ^{\frac{d+1}{2}}}\Bigg[\cosh(\mu)\csch^{d+1}(\mu)\times\\
        &\times\left(\frac{3\tanh^2(\mu)}{2(d-2)} \, _3F_2\left(1,1,\frac{5}{2};2,\frac{d+3}{2};\tanh^2(\mu)\right)+\frac{d+1}{d-2} \log (\sech(\mu))\right)\\
        &-\sech^d(\mu)\, _2F_1\left(\frac{d}{2},\frac{d+1}{2};\frac{d+3}{2};\tanh ^2(\mu)\right)\times\\
        &\times\Bigg(\frac{1}{d-1}\tanh^{-d+1}(\mu) \, _2F_1\left(\frac{1-d}{2},\frac{4-d}{2};\frac{3-d}{2};\tanh^2(\mu)\right)\\
        &-\frac{1}{4 \sqrt{\pi }}\Gamma \left(\frac{1-d}{2}\right) \Gamma \left(\frac{d-2}{2}\right)\Bigg)\Bigg].
    \end{aligned}
\end{equation}
As in the Fried-Yennie gauge case discussed in the main text, requiring now that at short distances the propagator is not more singular than in flat space fixes $c_0=0$ (and this automatically also ensures that the flat space limit behaviour of the propagator is correct). Demanding also that the propagator vanishes at large separation fixes $c=0$. Thus, we get
\begin{equation} \label{B_gen}
    \begin{aligned}
        B&=\left(d-(d-2)\xi\right)\frac{\Gamma\left(\frac{d+1}{2}\right)}{2\pi ^{\frac{d+1}{2}}}\Bigg[\cosh(\mu)\csch^{d+1}(\mu)\times\\
        &\times\left(\frac{3\tanh^2(\mu)}{2(d-2)} \, _3F_2\left(1,1,\frac{5}{2};2,\frac{d+3}{2};\tanh^2(\mu)\right)+\frac{d+1}{d-2} \log (\sech(\mu))\right)\\
        &-\sech^d(\mu)\, _2F_1\left(\frac{d}{2},\frac{d+1}{2};\frac{d+3}{2};\tanh ^2(\mu)\right)\times\\
        &\times\Bigg(\frac{1}{d-1}\tanh^{-d+1}(\mu) \, _2F_1\left(\frac{1-d}{2},\frac{4-d}{2};\frac{3-d}{2};\tanh^2(\mu)\right)\\
        &-\frac{1}{4 \sqrt{\pi }}\Gamma \left(\frac{1-d}{2}\right) \Gamma \left(\frac{d-2}{2}\right)\Bigg)\Bigg].
    \end{aligned}
\end{equation}
The position space propagator in covariant gauge and for general $\xi$ has also been computed in \cite{Ciccone:2024guw}. The expressions in \cite{Ciccone:2024guw} involve derivatives of the hypergeometric function with respect to the indices, which are difficult to deal with, even numerically. To the extent that we are able to check (numerically), the results here and in \cite{Ciccone:2024guw} are in agreement.  

\subsubsection*{Odd integer $d$}

Getting now back to the case when $d$ is an odd integer, the reason why this case is special can be seen from the integrand of $I_2$. Note that
\begin{equation}
    (1-t)^{\frac{d-4}{2}}=\sum_{k=0}^\infty \frac{(-1)^k \Gamma \left(\frac{d-2}{2}\right)}{\Gamma (k+1) \Gamma \left(\frac{d-2}{2}-k\right)}t^k,
\end{equation}
and hence $I_2$ can be rewritten as
\begin{equation}
    I_2=\frac12\int\sum_{k=0}^\infty \frac{(-1)^k \Gamma \left(\frac{d-2}{2}\right)}{\Gamma (k+1) \Gamma \left(\frac{d-2}{2}-k\right)}t^{k-\frac{d+1}{2}}.
\end{equation}
If $d$ is an odd integer, there will be a term proportional to $1/t$ in this sum which when integrated will give a $\log$ term. In the following we will assume that $d$ is an odd integer. The simplest way to deal with this is to separate the sum into three pieces as follows
\begin{equation}
    \begin{aligned}
        I_2&=\frac12\int\Bigg(\sum_{k=0}^{k=\frac{d-3}{2}}\frac{(-1)^k \Gamma \left(\frac{d-2}{2}\right)}{\Gamma (k+1) \Gamma \left(\frac{d-2}{2}-k\right)}t^{k-\frac{d+1}{2}}+\frac{(-1)^{\frac{d+1}{2}} \Gamma \left(\frac{d-2}{2}\right)}{4 \sqrt{\pi } \Gamma \left(\frac{d+1}{2}\right)t}\\
        &+\sum_{k=\frac{d+1}{2}}^{\infty}\frac{(-1)^k \Gamma \left(\frac{d-2}{2}\right)}{\Gamma (k+1) \Gamma \left(\frac{d-2}{2}-k\right)}t^{k-\frac{d+1}{2}}\Bigg)\\
        &=\sum_{k=0}^{\frac{d-3}{2}}\frac{(-1)^k \Gamma \left(\frac{d-2}{2}\right)}{\Gamma (k+1) \Gamma \left(\frac{d-2}{2}-k\right)}\frac{t^{k+1-\frac{d+1}{2}}}{2k+1-d}+\frac{(-1)^{\frac{d+1}{2}} \Gamma \left(\frac{d-2}{2}\right)}{4 \sqrt{\pi } \Gamma \left(\frac{d+1}{2}\right)}\log t\\
        &+\frac{3(-1)^{\frac{d+1}{2}} \Gamma \left(\frac{d-2}{2}\right)}{8 \sqrt{\pi } \Gamma \left(\frac{d+3}{2}\right)} t \, _3F_2\left(1,1,\frac{5}{2};2,\frac{d+3}{2};t\right)-\frac{(-1)^{\frac{d+1}{2}} \Gamma \left(\frac{d-2}{2}\right) \left(H_{\frac{d-1}{2}}-2+2\log 2\right)}{4 \sqrt{\pi } \Gamma
        \left(\frac{d+1}{2}\right)}
    \end{aligned}
\end{equation}
(where again the constant of integration has been chosen for later convenience). The complete solution in this case can be found as before by requiring that the propagator is not more singular than in flat space at short distances and that it vanishes at large distances, which once again will set the coefficients of the complementary function to zero. Thus, we get
\begin{equation} \label{B_odd}
    \begin{aligned}
        B&=\left(d-(d-2)\xi\right)\frac{\Gamma\left(\frac{d+1}{2}\right)}{2\pi ^{\frac{d+1}{2}}}\Bigg[\cosh(\mu)\csch^{d+1}(\mu)\times\\
        &\times\left(\frac{3\tanh^2(\mu)}{2(d-2)} \, _3F_2\left(1,1,\frac{5}{2};2,\frac{d+3}{2};\tanh^2(\mu)\right)+\frac{d+1}{d-2} \log (\sech(\mu))\right)\\
        &+\sech^d(\mu)\, _2F_1\left(\frac{d}{2},\frac{d+1}{2};\frac{d+3}{2};\tanh ^2(\mu)\right)\times\\
        &\times\Bigg(\sum_{k=0}^{\frac{d-3}{2}}\frac{(-1)^k \Gamma \left(\frac{d-2}{2}\right)}{\Gamma (k+1) \Gamma \left(\frac{d-2}{2}-k\right)}\frac{\tanh^{2k+1-d}(\mu)}{2k+1-d}+\frac{(-1)^{\frac{d+1}{2}} \Gamma \left(\frac{d-2}{2}\right)}{2 \sqrt{\pi } \Gamma \left(\frac{d+1}{2}\right)}\log \left(\tanh(\mu)\right)\\
        &+\frac{3(-1)^{\frac{d+1}{2}} \Gamma \left(\frac{d-2}{2}\right)}{8 \sqrt{\pi } \Gamma \left(\frac{d+3}{2}\right)} \tanh^2(\mu) \, _3F_2\left(1,1,\frac{5}{2};2,\frac{d+3}{2};\tanh^2(\mu)\right)\\
        &-\frac{(-1)^{\frac{d+1}{2}} \Gamma \left(\frac{d-2}{2}\right) \left(H_{\frac{d-1}{2}}-2+2\log 2\right)}{4 \sqrt{\pi } \Gamma
        \left(\frac{d+1}{2}\right)}\Bigg)\Bigg].
    \end{aligned}
\end{equation}

In both cases of generic values of $d$ and odd integer values of $d$, the solution for the propagator is now complete as $A$ can be determined using equation (\ref{A in terms of B}), namely
\begin{equation}
    A=-\xi\frac{ \Gamma \left(\frac{d+1}{2}\right)}{2 d \pi ^{\frac{d+1}{2}}}\csch^{d-2}(\mu)+\frac{1}{d}\sinh ^2(\mu) B'+ \sinh (\mu) \cosh (\mu)B.
\end{equation}
Plugging in $d=3,4,5,6$ in these general formulas recovers the specific solutions listed above. It is straightforward to see that in the Fried-Yennie gauge $\xi=d/(d-2)$ the above expressions for $B$ are zero and the expression for $A$ simplifies to (\ref{form factors in FY gauge}). We have also (partially) checked numerically that in the limit where $d$ becomes an odd integer, the general expression \eqref{B_gen} limits to \eqref{B_odd}.
\section{Special case: $d=2$}\label{app: d=2}

The case when $d=2$ is special. Maxwell theory in $d+1=3$ dimensions is special both classically and quantum mechanically, already in flat space. At the classical level, the Coulomb potential exhibits a logarithmic behaviour. Quantum mechanically, the Maxwell theory is super-renormalizable. While this implies good UV behaviour, it also signals strong IR divergences. Indeed, as we will see momentarily, the propagator exhibits logarithmic behaviour at large separation (and thus lack of clustering), instead of the power follow-off that we saw in higher dimensions. In more detail, in Poincare coordinates the two independent solutions of the classical equations of motion of the Maxwell theory in $d$ dimensions behave as $z^0$ and $z^{d-2}$ close to the boundary. When $d=2$ one of these turns into a $\log$, which is divergent as $z\to0$.  This issue is manifested in the general-$d$ expressions for the propagator by $1/(d-2)$ poles. The case of three bulk dimensions is also special in that there is a lower-dimension term one may add to the action, the Chern–Simons term. In explicit examples that come from string theory, the low-energy effective action typically contains a Chern-Simons term and the relevant bulk-to-bulk propagator would be the one that includes the Chern-Simons term. Such a computation appeared recently in \cite{Bhattacharya:2025udq}.  

To work out the propagators corresponding to the Maxwell theory we need to set $d=2$ from the beginning. Following the same steps as in the main text, one then arrives at the following propagators in the covariant gauge and in position space,  
\begin{align}
    G_{\rm ghost}(\mu)&=-\frac{1}{4\pi}(\coth(\mu)-1),\\
    A(\mu)&=\frac{1}{8\pi}\Big(-2-\xi+2(-2\pi c+\mu)\coth(\mu)+(4\pi c-\mu)\mu\csch^2(\mu)\Big),\\
    B(\mu)&=\frac{1}{4\pi}\Big(4\pi c-2\mu+\mu(-4\pi c+\mu)\coth(\mu)\Big)\csch^2(\mu).
\end{align}
As alluded above, the photon propagator diverges at large separation: as one takes one of the points to the boundary there is a $\log$ divergence that matches the one from the solutions of the classical equations of motion. The propagator depends on the arbitrary constant $c$, which is related to the scale of the $\log$ coefficient. We have checked that this propagator (with $\xi=c=0$) agrees with the even part of the propagator derived in \cite{Bhattacharya:2025udq}, when the Chern-Simons coupling is sent to zero. 

One may similarly work out the propagator in other gauges. For example, in axial gauge the photon propagator terms involving $z^{d-2}/(d-2)$ get replaced by $\log$s and once again there is a constant associated with the scale of the $\log$s. The Coulomb gauge photon propagator remains unchanged.

\bibliography{bibfile}{}
\bibliographystyle{utcaps}

\end{document}